\documentclass[prd,onecolumn]{revtex4}
\usepackage{dcolumn}
\usepackage{multirow}
\usepackage{graphicx}
\usepackage{amssymb}
\usepackage{bm}
\usepackage{hyperref}
\usepackage{epstopdf}
\usepackage{color}
\usepackage{mathrsfs}
\usepackage{amsmath,amssymb,amsthm}

\begin{document}
\title{Observational constraints and diagnostics for time-dependent dark energy models}
\author{Deng Wang}
\email{Cstar@mail.nankai.edu.cn}
\affiliation{Theoretical Physics Division, Chern Institute of Mathematics, Nankai University,
Tianjin 300071, China}
\author{Xin-he Meng}
\email{xhm@nankai.edu.cn}
\affiliation{{Department of Physics, Nankai University, Tianjin 300071, P.R.China}\\
{State Key Lab of Theoretical Physics,
Institute of Theoretical Physics, CAS, Beijing 100080, P.R.China}}
\begin{abstract}
In this paper, we constrain four time-dependent dark energy (TDDE) models by using the Type Ia supernovae (SNe Ia), baryonic acoustic oscillations (BAO), observational Hubble parameter (OHD) data-sets as well as the single data point from the newest event GW150914. Subsequently, adopting the best fitting values of the model parameters, we apply the original statefinder, statefinder hierarchy, the growth rate of matter perturbations and $Om(z)$ diagnostics to distinguish the TDDE scenarios and the $\Lambda$CDM scenario from each other. We discover that all the TDDE models and $\Lambda$CDM model can be distinguished better at the present epoch by using the statefinder hierarchy than using the original statefinder, the growth rate of matter perturbations and $Om(z)$ diagnostics, especially, in the planes of $\{S_3^{(1)},S_4^{(1)}\}$, $\{S_3^{(2)},S_4^{(2)}\}$, $\{S_5^{(1)},S_5^{(2)}\}$ and $\{S_4^{(2)},S_5^{(2)}\}$.

\end{abstract}
\maketitle
\section{Introduction}
Modern astrophysical observations such as the measurements of Type Ia supernovae (SNe Ia), the cosmic microwave background (CMB) anisotropy, the baryonic acoustic oscillations (BAO) measurement from the Sloan Digital Sky Survey (SDSS) and so on, have confirmed our universe is undergoing a phase of accelerated expansion at the present epoch \cite{1,2,3}. In the past few years, cosmologists have introduced an additional component in the matter and energy sector, named dark energy, to explain the accelerated mechanism. The simplest and most attractive candidate of dark energy is the so-called $\Lambda$CDM model \cite{4}, which has been proved to be very successful in describing many aspects of the observed universe. The newest results of Planck 2015 for the $\Lambda$CDM cosmology have shown that all the conclusions, as in its 2013 analysis, are in good agreement with the JLA sample of SNe Ia and BAO data-sets \cite{5}. However, besides the observed $H(z)$ anomaly, one of the other anomalies the amplitude of fluctuation spectrum is still found to be higher than deduced from the analysis of weak gravitational lensing and rich cluster counts. At the same time, the authors also show that the tensions can not be resolved with some simple modifications of the $\Lambda$CDM model. In addition, this model also faces two fatal detects, i.e., the `` coincidence '' problem and the `` fine-tuning '' problem \cite{4}. The former implies why the amounts of the dark matter and dark energy are at the same order today since the scaling behavior of the energy densities are substantially different during the evolution of the universe by global fitting, while the latter indicates that the measured energy density of the vacuum is much smaller than the theoretical prediction value, which is the so-called 120-orders-of-magnitude discrepancy that makes the vacuum explanation so suspicious. Thus, the the actual nature and cosmological origin of dark energy might not be the cosmological constant $\Lambda$ in the standard cosmological model. Based on this concern, in recent years, theorists have proposed many alternatives to explain the dark energy phenomenon including phantom \cite{6}, quintessence \cite{7,8,9,10,11,12,13,14,15}, quintom \cite{16}, bulk viscosity \cite{23,24,25,26,27,28,28.1}, generalized Chaplygin gas (GCG) \cite{29,30}, modified Chaplygin gas (MCG) \cite{31,32}, superfluid Chaplygin gas (SCG) \cite{33,34,35}, decaying vacuum \cite{36}, time-dependent dark energy (TDDE) \cite{37,38,39,40,41,42,43,43.1}, holographic dark energy (HDE) \cite{44,45,46,47}, Ricci dark energy (RDE) \cite{48,49,50,51}, holographic tachyon model \cite{52,53}, f(R) gravity \cite{54,55,56,57}, scalar-tensor theories of gravity \cite{58,59,60,61,62,63,64}, Gauss-Bonnet gravity \cite{A,B,C,D}, Einstein-Aether gravity \cite{65,66}, braneworld models \cite{67,68,69,70}, etc.

Since so many dark energy models have been proposed, it becomes substantially important and constructive to discriminate them from the $\Lambda$CDM model and one from the other in order to find better scenarios. As is well known, one can think of the expansion rate of the universe as the Hubble parameter $H=\dot{a}/a$, where $a$ is the scale factor, while the rate of the universe acceleration can be explained by the deceleration parameter
\begin{equation}
q=-\frac{\ddot{a}}{aH^2}=-\frac{a\ddot{a}}{\dot{a}^2}.
\end{equation}
Nonetheless, the Hubble parameter $H$ and the deceleration parameter $q$ cannot differentiate various dark energy models more accurately since all the models will lead to the same result, namely, $\ddot{a}>0$ and $H>0$ or $q<0$. Moreover, the gradually mounting observation data-sets with higher precision and more advanced statistical methods force us to invoke some newer and more effective quantities to surpass the two original quantities. Therefore, naturally, an interesting and appealing question occurs: how can one differentiate various kinds of dark energy cosmological models more explicitly and efficiently ? In order to solve this problem, recently, a new geometrical diagnostic called statefinder is proposed in \cite{71,72}, which involves the third derivative of the scale factor $a$. The statefinder $\{r,s\}$ can be defined as follows:
\begin{equation}
r=\frac{\dddot{a}}{aH^3},\qquad s=\frac{r-1}{3(q-1/2)}.
\end{equation}
As usual, one can plot the corresponding trajectories for various dark energy models in the $r-s$ plane in order to investigate qualitatively the different behaviors. For the $\Lambda$CDM model, the statefinder pair corresponds to the fixed point $\{1,0\}$, which can be regarded as a basic point to measure the distance of any given dark energy model from the standard cosmological model. Recently, the statefinder has been used to discriminate a great deal of dark energy models, for instance, quintessence \cite{73,74,75}, quintom \cite{76}, parametrization models for effective pressure \cite{77}, purely kinetic k-essence (PKK) model \cite{78}, GCG \cite{79,80,81}, HDE \cite{82,83}, RDE \cite{84}, agegraphic dark energy (ADE) model \cite{85}, spatial Ricci dark energy model (SRDE) \cite{86}, Dvali-Gabadadze-Porrati (DGP) gravity \cite{87,88}, Galileon modified gravity \cite{89}, HDE in the DGP braneworld \cite{90}, etc.

Apart from the statefinder, another useful diagnostic, namely, the $Om(z)$ diagnostic \cite{91}, has been applied into discriminating different dark energy models. The $Om(z)$ method can be constructed from the Hubble parameter $H(z)$ and it remains invariable at different stages of the universe for the $\Lambda$CDM model. Therefore, this diagnostic gives a simple null test to discriminate the $\Lambda$CDM scenario from the evolving dark energy models, since the values of $Om(z)$ for various cosmological models are the functions of the redshift. Furthermore, according to Ref. \cite{91}, one can also obtain the conclusion that $Om(z)$ does not use any information about the evolution of inhomogeneities in the Friedmann-Robertson-Walker (FRW) background, and can not discriminate between large and small values of the cosmological constant unless the value of matter density has been independently known. In addition, it is worth mentioning that $Om(z)$ is a relatively sketchy diagnostic which just depends on a knowledge of the Hubble parameter and can be determined well with currently observational data-sets. Lately, the $Om(z)$ diagnostic has been adopted to distinguish various dark energy models from the so-called $\Lambda$CDM model, for instance, phantom \cite{91}, quintessence \cite{91}, parametrization models for effective pressure \cite{77}, PKK \cite{78}, HDE \cite{82}, and SRDE \cite{86}. An attractive extension of the $Om(z)$ diagnostic, named $Om3(z)$ diagnostic \cite{92}, which is reconstructed from the SNe Ia and BAO date-sets, has provided a powerful null diagnostic for the $\Lambda$CDM scenario from the other cosmological scenarios. The $Om3(z)$ diagnostic, acting as a three-point diagnostic tool of the dark energy cosmological models, is very closely related to the $Om(z)$ diagnostic, and follows the same basic principles. However, the $Om3(z)$ method has a unique advantage, i.e., its value does not depend on either the distance to the last scattering surface or the present-day values of the Hubble parameter $H(z)$ and the matter density parameter $\Omega_{m}$. Hence, the uncertainties of these observational quantities, will not have an effect on the reconstruction of the $Om3(z)$ diagnostic. For this method, it is necessary to point out that we need the more accurate data such as the BigBoss experiment, in order to put more tighter constraints on the $Om3(z)$ diagnostic, since the uncertainties of the presently available BAO data is substantially large.

Since the original statefinder parameters are only related to the third derivative of scale factor, one may not distinguish well various dark energy models from each other. Therefore, Arabsalmani et al. propose an extended null diagnostic for the base cosmology scenario, namely, the statefinder hierarchy, which contains the higher derivatives of scale factor. They demonstrate that, for the base cosmology scenario, all the members of the statefinder hierarchy can be expressed in terms of the elementary functions of the deceleration parameter $q$, consequently and of the matter density parameter $\Omega_{m}$. This feature can be employed to discriminate better the evolving dark energy models from the $\Lambda$CDM model, since the statefinder hierarchy also remain pegged at one fixed point as the statefinder diagnostic and $Om(z)$ diagnostic during the evolution of the universe. For instance, in paper \cite{78}, the authors have demonstrated that one can not discriminate the PKK model from the $\Lambda$CDM model in terms of 68.3\% confidence level through adopting the $Om(z)$ method and statefinder pair $\{r,s\}$. Subsequently, Li et al. \cite{93} exhibit that they can distinguish well the PKK model from the $\Lambda$CDM model as well as other dark energy models by adopting the statefinder hierarchy and the growth rate of matter perturbations. Thus, these two methods can act as the starting point of our work. In this situation, we would like to use the $Om(z)$ diagnostic, the statefinder hierarchy and the growth rate of matter perturbations to distinguish four different time-dependent dark energy models from the $\Lambda$CDM model, and one from the other, in order to find better ones as the following examples.

This paper is organized as follows: In the next section, we will make a brief review about these four TDDE models. In Section III, we would like to constrain the models by using the SNe Ia, BAO, OHD data-sets as well as the single data point from the newest event GW150914 \cite{94}. In Section IV, we briefly review the statefinder hierarchy, the growth rate of matter perturbations and the $Om(z)$ diagnostic. In Section V, we discriminate the four models by using the aforementioned three diagnostics. In the final section, the concluding remarks are presented.
\section{The Four Time-dependent dark energy models}
Parameterization has been applied to analyze various kinds of astronomical data-sets, which is proved to be an very impactful and useful tool towards a more complete description of dark energy modelling.
Generally speaking, phenomenologically, one can explore some possible time-dependent parameterizations to describe the dark energy equation of state parameter $\omega(z)$. Furthermore,  by Taylor-expanding $\omega(z)$, one can obtain the following parameterizations formalism
\begin{equation}
\omega(z)=\sum_{n=0}\omega_nx_n(z),
\end{equation}
where $\omega_n$ denote the parameters to be determined by astrophysical observations and $x_n(z)$ the functions of the redshift $z$. Obviously, one can get different parameterized dark energy models by making some subtle choices of the functions $x_n(z)$.

In this section, we will make a brief introduction about the four time-dependent dark energy models. In addition, it is worth noting that we will neglect the radiation contribution at low redshifts and consider the flat FRW background spacetime throughout the context.
\subsection{Model 1}
The first parameterization model was firstly proposed in \cite{37} and can be expressed as
\begin{equation}
\omega(z)=\omega_0+\omega_1z,
\end{equation}
where $\omega_0$ denotes the present-day value of the equation of state parameter, and $\omega_1$ a free parameter to be determined by observations. This model was firstly constrained by Cooray et al. \cite{95} through adopting the SNe Ia data-set, gravitational lensing statistics and global clusters ages. At the same time, Goliath et al. \cite{39} also studied the limit consequences of this model from the SNe Ia experiments. This parameterization is a good fit for low redshifts , and is exact for models where the equation of state changes slowly or is a constant. Nonetheless, this model exhibit a problematic behavior for high redshifts, such as failing to explain the age estimations of high-$z$ objects since it just predicts substantially small ages at $z\geqslant3$. For conveniences of following constraints, the dimensionless Hubble parameter including the dust matter and dark energy is given by
\begin{equation}
E(z)=\frac{H(z)}{H_0}=[\Omega_{m0}(1+z)^3+(1-\Omega_{m0})(1+z)^{3(1+\omega_0-\omega_1)}e^{3\omega_1z}]^{1/2},
\end{equation}
where the parameters $\omega_0$, $\omega_1$ and $\Omega_{m0}$ will be determined by the following astrophysical observations.
\subsection{Model 2}
The second parameterization model was introduced by Efstathiou \cite{40}, which is aimed at adjust some quintessence models at $z\lesssim4$. The author discovered that, for a wide class of potentials related to the dynamical scalar field models, the evolutional behavior of $\omega(z)$ at $z\lesssim4$ can be well approximated by the following parameterization formalism
\begin{equation}
\omega(z)=\omega_0-\omega_2\ln(1+z),
\end{equation}
where $\omega_2$ denotes a free parameter to be determined by astronomical observations. Subsequently, the corresponding dimensionless Hubble parameter can be expressed as
\begin{equation}
E(z)=[\Omega_{m0}(1+z)^3+(1-\Omega_{m0})(1+z)^{3(1+\omega_0)}e^{-\frac{3\omega_2[\ln(1+z)]^2}{2}}]^{1/2},
\end{equation}
where the parameters $\omega_0$, $\omega_2$ and $\Omega_{m0}$ will be determined by the following astrophysical observations.
\subsection{Model 3}
The third parameterization scenario is the so-called CPL (Chevallier-Polarski-Linder) parameterization \cite{41}, which is intended to solve the problematic behavior at high redshifts. Furthermore, this scenario is an excellent fit for a number of theoretically conceivable scalar field potential, give a good explanation for small deviations from the phantom barrier $\omega=-1$ (see also \cite{96}). At the same time, $\omega(z)$ is a well behaved function at $z\gg1$, and recovers the linear behavior at low redshifts. Therefore, it is worth investigating this scenario further and it can be expressed in the following manner
\begin{equation}
\omega(z)=\omega_0+\omega_3(\frac{z}{1+z}),
\end{equation}
where $\omega_3$ is also a free parameter to be determined by astrophysical observations. Subsequently, the corresponding dimensionless Hubble parameter can be written as
\begin{equation}
E(z)=[\Omega_{m0}(1+z)^3+(1-\Omega_{m0})(1+z)^{3(1+\omega_0+\omega_3)}e^{-\frac{3\omega_3z}{1+z}}]^{1/2},
\end{equation}
similarly, where the parameters $\omega_0$, $\omega_3$ and $\Omega_{m0}$ should be determined by the following astrophysical observations.
\subsection{Model 4}
The mentioned-above three parameterizations can not be reconstructed from the scalar field dynamics since they are not bounded functions, namely, the equation of state parameters are all divergent functions of the redshift $z$, which lie in the range $z\in[-1,\infty)$. Since the dark energy phenomenon occurs being not far away from the present epoch, the above three models can provide substantially good approximations and exhibit a quintom-like behavior to describe it, when $z$ is relatively finite. Even so, one can query the information of the aforementioned parameterizations has been compromised so as to we can not identify. In this concern, Barboza et al. \cite{43.1} proposed a new parameterization model which is aimed at extending the range of applicability of the dark energy equation of state, and avoid the singularities and uncertainties contained in three mentioned-above scenarios. In addition, one can obtain this new scenario from the scalar field dynamics. The last parameterization model could be expressed in the following manner:
\begin{equation}
\omega(z)=\omega_0+\omega_4\frac{z(1+z)}{1+z^2},
\end{equation}
where $\omega_4$ is a free parameter to be ensured by observations. Subsequently, the corresponding dimensionless Hubble parameter can be expressed as
\begin{equation}
E(z)=[\Omega_{m0}(1+z)^3+(1-\Omega_{m0})(1+z)^{3(1+\omega_0)}(1+z^2)^{\frac{3\omega_4}{2}}]^{1/2},
\end{equation}
as before, where the parameters $\omega_0$, $\omega_4$ and $\Omega_{m0}$ will be ensured by the following astronomical observations.
\section{Observational constraints}
\subsection{Type Ia Supernovae Observations}
In this situation, we adopt the Union 2.1 SNe Ia data-sets without systematic errors for fitting, which covers the redshift range $z\in[0.015,1.4]$. The theoretical distance modulus for a supernovae at redshift $z$, given a set of model parameters $K$, is
\begin{equation}
\mu_{t}(z;K)=m-M=5\lg d_L+25,
\end{equation}
where $m$ denotes the apparent magnitude, $M$ the absolute magnitude and $d_L$ the luminosity distance in units of megaparsecs,
\begin{equation}
d_L(z;K)=(1+z)\int^z_0\frac{dz'}{E(z';K)},
\end{equation}
where $E(z;K)$ represents the dimensionless Hubble parameter for a concrete dark energy model, such as Eqs. (5), (7), (9) and (11). Subsequently, we will calculate the best fitting values for the model parameters $K$ by performing the so-called $\mathcal{\chi}^2$ statistics, namely,
\begin{equation}
\chi^2_S=\sum^{580}_{i=1}\frac{[\mu^i_t(z;K)-\mu^i_o(z)]^2}{\sigma^2_i},
\end{equation}
where $\mu^i_o(z)$ and $\sigma_i$ denote the observed value and the corresponding $1\sigma$ error of the distance modulus, respectively, for a given supernovae at $z_i$.
\subsection{Baryonic Acoustic Oscillations}
As is well known, the spatial two-point correlation function of the density of baryons has a peak, namely, the BAO peak, at a comoving scale $r_s$ which is proved to be about 150 Mpc. Since the baryons on these scales are non-relativistic after short recombination, the location of the peak in the comoving frame would not change. Hence, the location of the peak provides a general ruler, with a constant comoving scale at distinguishable redshifts during almost the whole cosmic history. Furthermore, we adopt the BAO data-sets which can be found in \cite{97}, and use the parameter $\mathcal{A}$ to measure the BAO peak in the distribution of the SDSS luminous red galaxies. Then, the parameter $\mathcal{A}$ can be defined as
\begin{equation}
\mathcal{A}=\sqrt{\Omega_{m0}}E(z_i)^{-\frac{1}{3}}[\frac{1}{z_i}\int^{z_i}_0\frac{dz'}{E(z')}]^{\frac{2}{3}}.
\end{equation}
The corresponding $\mathcal{\chi}^2$ for the BAO measurements is
\begin{equation}
\chi^2_{B}=\sum^6_{i=1}[\frac{\mathcal{A}_{o}(z_i)-\mathcal{A}_{t}(z_i;K)}{\sigma_{\mathcal{A}}}]^2,
\end{equation}
where the parameters $\mathcal{A}_{o}$ and $\mathcal{A}_{t}$ denotes the observed value and the theoretical value, respectively.
\subsection{Observational Hubble Parameter}
In our combined analysis we will use 29 determinations of the Hubble expansion parameter $H(z)$ as a function of the redshift $z$. These determinations are obtained by two basic methods, i.e., `` differential age method radial BAO method '' and `` radial BAO method ''. More useful information can be found in \cite{98,99}. Comparing with the two above observations, the $H(z)$ data-sets can directly reflect the expansion rate of the universe and there is no need to integrate over the redshift $z$ so as to drop out some useful information when constraining a concrete model. To perform the interesting test one can minimize the following quantity:
\begin{equation}
\chi^2_{H}=\sum^{29}_{i=1}[\frac{H_0E(z_i)-H_{obs}(z_i)}{\sigma_i}]^2,
\end{equation}
where $H_{obs}(z_i)$ is the observed value of the Hubble expansion rate at a given $z_i$.

\subsection{The Gravitational Wave}
On September 14, 2015 at 09:50:45 UTC the two detectors of the Laser Interferometer Gravitational-Wave Observatory observed a transient gravitational wave signal from a black hole-black hole binary (BHBH) inspiral \cite{94}. This gravitational source lies at the luminosity distance of $410^{+160}_{-180}$ Mpc corresponding to the redshift $z=0.09^{+0.03}_{-0.04}$. In this situation, we would like to use this single data point of the gravitational wave to constrain the TDDE models as well. Although the quality of the data is not very good, we believe that the forthcoming gravitational-wave data-sets will provide a new and powerful window for new physics. Conveniently, we add this data point into the SNe Ia data-sets (580 data points) after calculating out the correspondingly observational distance modulus $38.0639^{+0.7155}_{-1.2553}$. In the following context, for simplicity, we will denote the statistical contribution from the gravitational wave data as $\chi^2_G$.

Subsequently, in the first place, we would compute the joint constraints from SNe Ia, BAO and OHD data-sets, and the corresponding $\chi^2_1$ can be defined as
\begin{equation}
\chi^2_{1}={\chi}^2_{S}+\chi^2_{B}+{\chi}^2_{H}.
\end{equation}

In the second place, we shall calculate the joint constraints from SNe Ia, BAO, OHD and the gravitational wave data-sets. The corresponding $\chi^2_2$ can be defined as
\begin{equation}
\chi^2_{2}={\chi}^2_{S}+\chi^2_{B}+{\chi}^2_{H}+\chi^2_G.
\end{equation}
The minimum values of the derived $\chi^2_{1}$ and the best fitting values of the model parameters constrained by SNe Ia, BAO and OHD data-sets, are listed in Table. \ref{tab1}. At the same time, The minimum values of the derived $\chi^2_{2}$ and the best fitting values of the model parameters constrained by SNe Ia, BAO, OHD as well as the gravitational wave data-sets, are listed in Table. \ref{tab2}.
\begin{table}[h!]
\begin{tabular}{ccccccc}
\hline
                      &model 1           & model 2       &model 3           &model 4\\
\hline
$\chi^2_{min}$        & $580.02$       &$580.024$      &$579.992$        &$580.047$\\
$\Omega_{m0}$         & $0.285497$      &$0.286945$      & $0.286345$     & $0.286367$\\
$\omega_0$            & $-1.00602$       &$-0.977002$      & $-0.981093$     & $-0.995272$ \\
$\omega_i$            &$-0.138015$        &$0.314506$      &$-0.325841$      &$-0.162797$\\
\hline
\end{tabular}
\caption{The best fitting values of the model parameters ($\Omega_{m0}$, $\omega_0$, $\omega_i$) in the TDDE models by using the combined constraints from the SNe Ia, BAO, and OHD data-sets, where $\omega_i$ denotes $\omega_1$, $\omega_2$, $\omega_3$ and $\omega_4$, respectively.}
\label{tab1}
\end{table}
\begin{table}[h!]
\begin{tabular}{ccccccc}
\hline
                      &model 1           & model 2       &model 3           &model 4\\
\hline
$\chi^2_{min}$        & $580.107$       &$580.191$      &$580.098$        &$580.16$\\
$\Omega_{m0}$         & $0.285777$      &$0.287902$      & $0.287025$     & $0.285128$\\
$\omega_0$            & $-1.00515$       &$-0.959598$      & $-0.96645$     & $-1.01115$ \\
$\omega_i$            &$-0.132859$        &$0.419564$      &$-0.429445$      &$-0.0823002$\\
\hline
\end{tabular}
\caption{The best fitting values of the model parameters ($\Omega_{m0}$, $\omega_0$, $\omega_i$) in the TDDE models by using the combined constraints from the SNe Ia, BAO, OHD data-sets as well as the single gravitational-wave data point, where $\omega_i$ denotes $\omega_1$, $\omega_2$, $\omega_3$ and $\omega_4$, respectively.}
\label{tab2}
\end{table}

\begin{figure}
\centering
\includegraphics[scale=0.7]{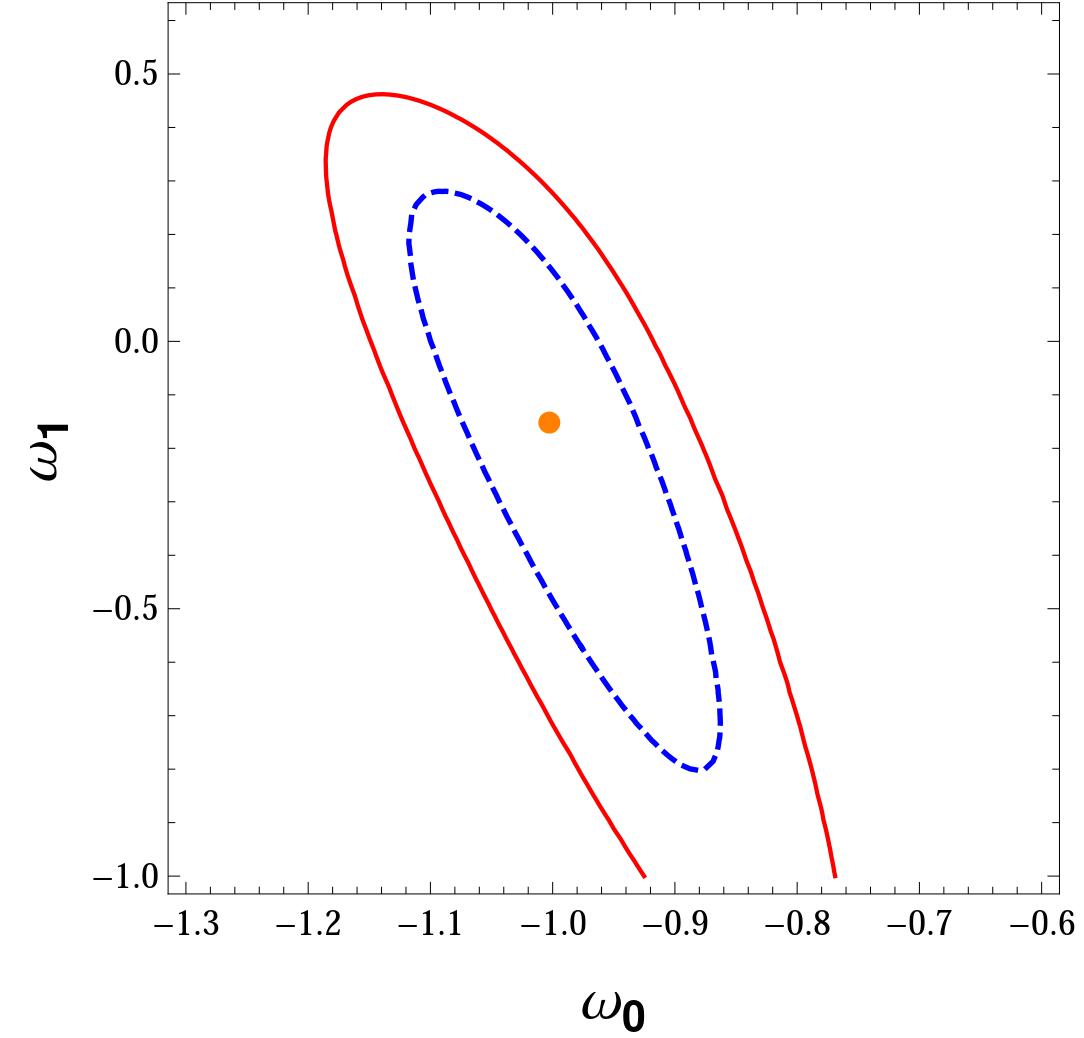}
\includegraphics[scale=0.7]{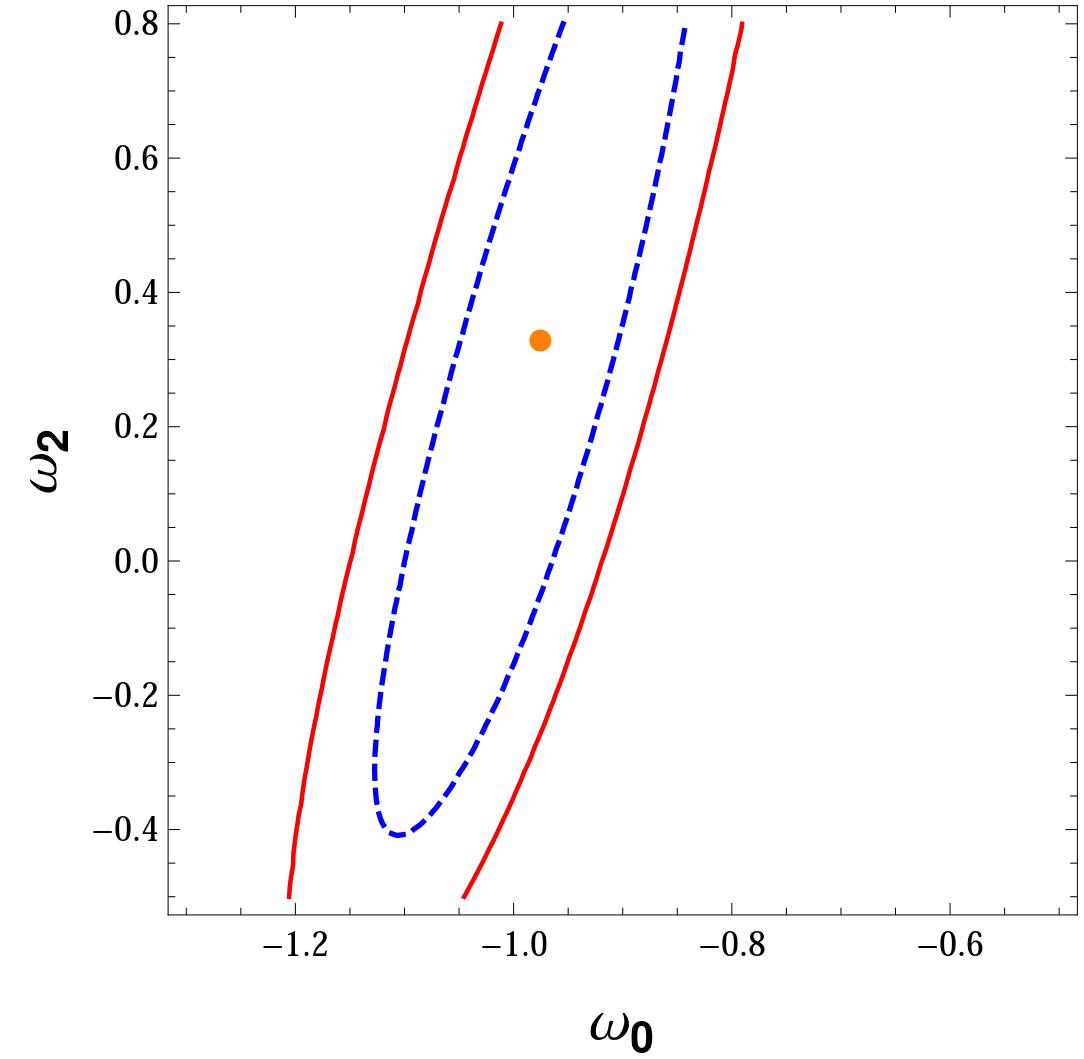}
\includegraphics[scale=0.7]{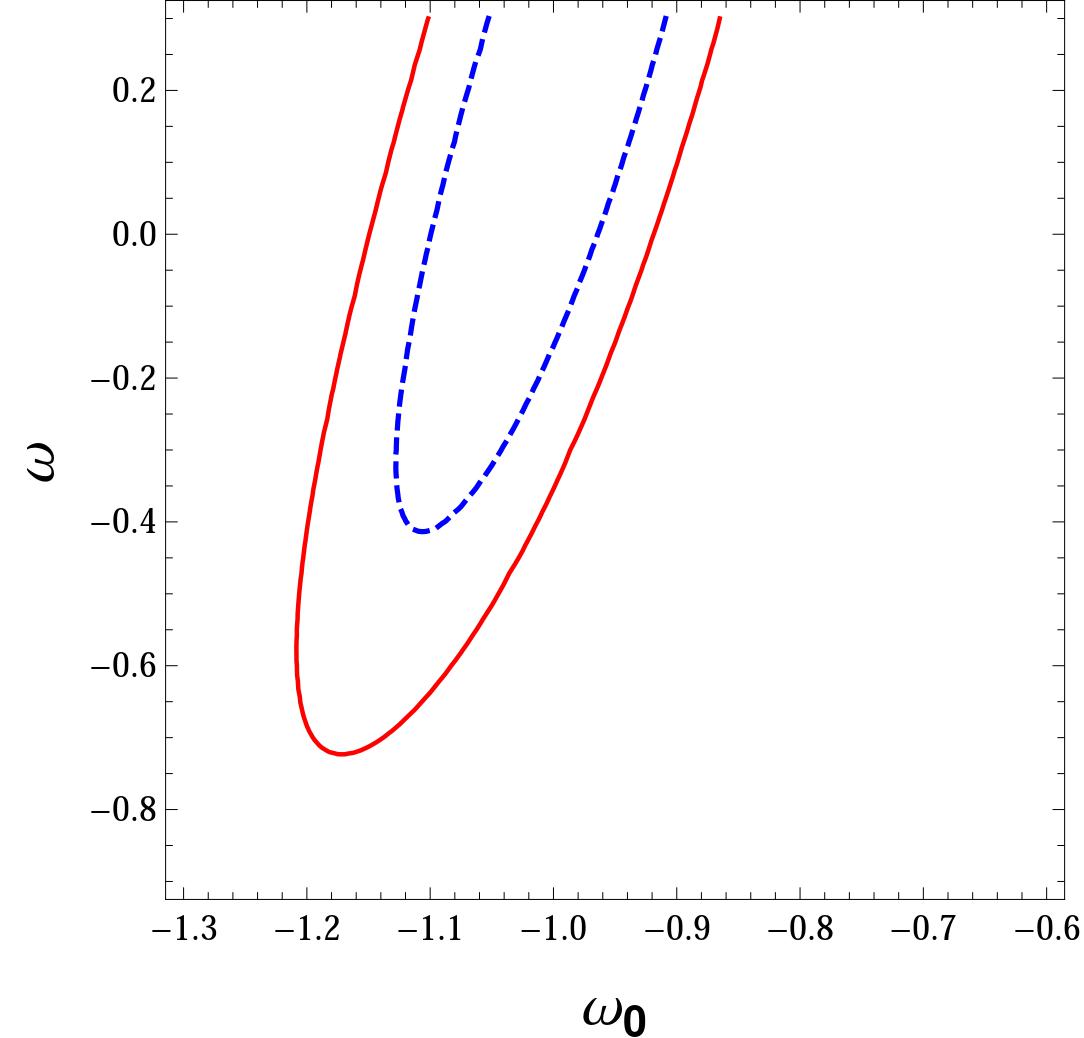}
\includegraphics[scale=0.7]{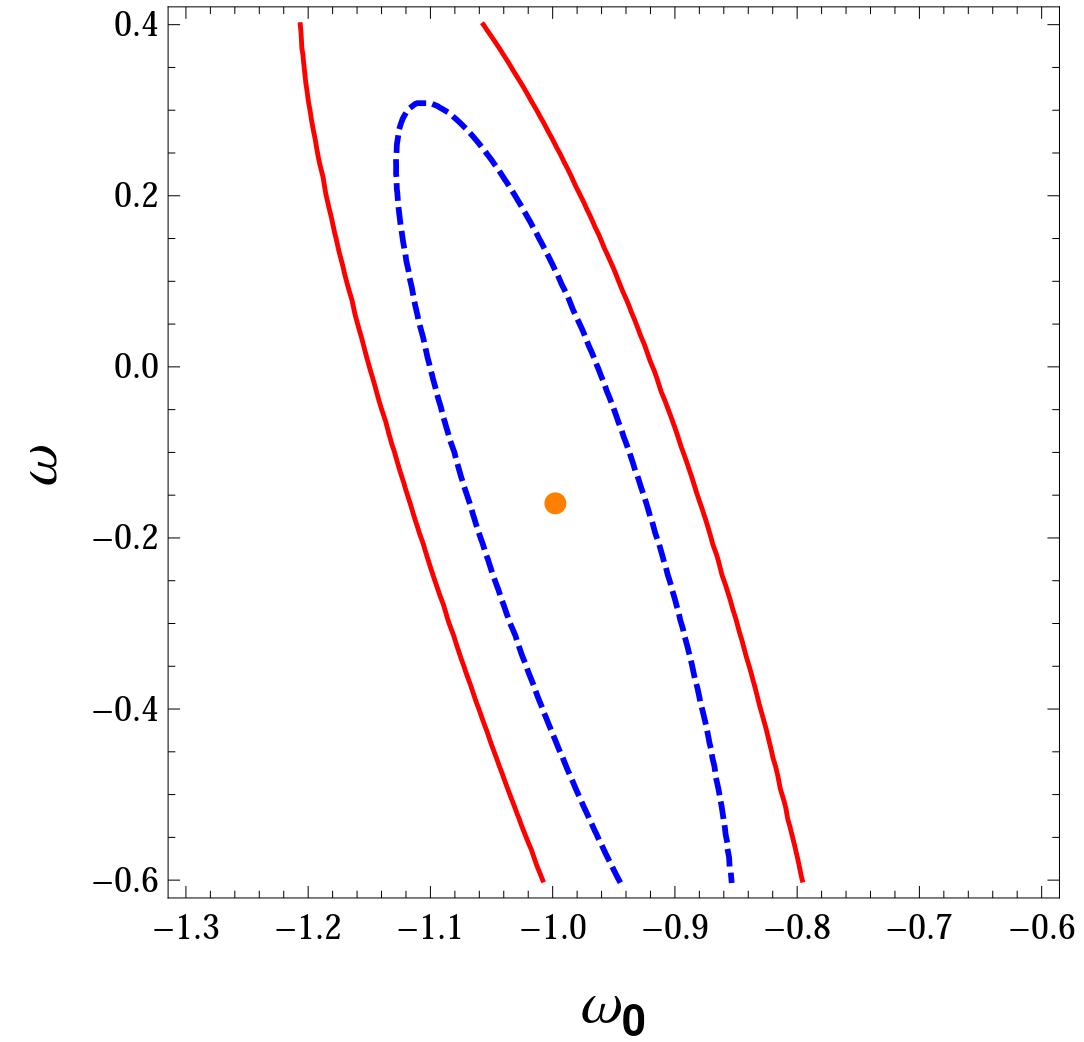}
\caption{1$\sigma$ and 2$\sigma$ confidence ranges for parameter pair ($\omega_0$, $\omega_i$) of the TDDE models, constrained by SNe Ia, BAO and OHD data-sets. The upper left panel, the upper right panel, the lower left panel and the lower right panel correspond to the likelihood distributions of model 1, model 2, model 3 and model 4. The best fitting value is shown as a dot for different models.}\label{0}
\end{figure}

In Figure. \ref{0}, we just perform the likelihood distributions of the parameters ($\omega_0$, $\omega_i$) for the first joint constraints $\chi^2_{1}$ from the SNe Ia, BAO and OHD data-sets, where $\omega_i$ denotes $\omega_1$, $\omega_2$, $\omega_3$ and $\omega_4$, respectively. In the following content, we still apply the best fitting values of the model parameters from the first joint constraints $\chi^2_{1}$ into distinguishing the TDDE models and the $\Lambda$CDM model from each other.

\section{The three diagnostics}
\subsection{The statefinder hierarchy}
As mentioned above, the statefinder hierarchy contains higher derivatives of the scale factor $d^na/dt^n$, $n\geqslant2$ so that it may discriminate the dark energy models better. According to \cite{92}, the scale factor can be obtained in the following manner by Taylor expansion:
\begin{equation}
\frac{a(t)}{a_0}=1+\sum\limits_{n=1}^\infty\frac{A_n(t_0)}{n!}[H_0(t-t_0)]^n \label{eqs-3},
\end{equation}
where
\begin{equation}
A_n=\frac{a^{(n)}}{aH^n} \label{eqs-4},
\end{equation}
where $a^{(n)}=d^na/dt^n$. It is worth noting that a number of letters of the alphabet have been adopted to represent different derivatives of the scale factor $a$. To be more precise, historically, $q=-A_2$ denotes the deceleration or acceleration parameter, $A_3$ the statefinder $r$ or the jerk $j$ \cite{100}, $A_4$ the snap \cite{100,101,102,103} and $A_5$ the lerk \cite{100,101,102,103}. Obviously, for the base cosmology in the spatially flat FRW universe, one can easily obtain
\begin{eqnarray}
  A_2&=&1-\frac{3}{2}\Omega_m \label{eqs-5}, \\
  A_3&=&1 \label{eqs-6},  \\
  A_4&=&1-\frac{3^2}{2}\Omega_m \label{eqs-7},  \\
  A_5&=&1+3\Omega_m+\frac{3^3}{2}\Omega^2_m  \label{eqs-8},\qquad ...,
\end{eqnarray}
where $\Omega_m=\Omega_{m0}(1+z)^3/E^2(z)$ and $\Omega_m=2(1+q)/3$ for the base cosmology. The statefinder hierarchy $S_n$ can be defined as
\begin{eqnarray}
  S_2&=&A_2+\frac{3}{2}\Omega_m \label{eqs-9}, \\
  S_3&=&A_3 \label{eqs-10},   \\
  S_4&=&A_4+\frac{3^2}{2}\Omega_m \label{eqs-11},   \\
  S_5&=&A_5-3\Omega_m-\frac{3^3}{2}\Omega^2_m  \label{eqs-12},\qquad ....
\end{eqnarray}
It is not difficult to verify that, for the base cosmology, the statefinder hierarchy $S_n$ can be rewritten as[23]
\begin{equation}
S_n\mid_{\Lambda CDM}=1 \label{eqs-12'}.
\end{equation}
It is noteworthy that the above equations just define a good mull diagnostic for the base cosmology, since the equalities will be violated by other dark energy models. Furthermore, when $n\geqslant3$, one can define a series of statefinders as follows:
\begin{eqnarray}
  S^{(1)}_3&=&S_3 \label{eqs-13}, \\
  S^{(1)}_4&=&A_4+3(1+q) \label{eqs-14},   \\
  S^{(1)}_5&=&A_5-2(4+3q)(1+q)  \label{eqs-15},\qquad....
\end{eqnarray}
The series of statefinders have the same property with $S_n$, namely, remaining pegged at unity during the evolution of the universe for the $\Lambda$CDM cosmology:
\begin{equation}
S^{(1)}_n\mid_{\Lambda CDM}=1 \label{eqs-16}.
\end{equation}
Therefore, one can obtain an interesting and important property, i.e., $\{S_n,S_n^{(1)}\}\mid_{\Lambda CDM}=1$. Similarly, other dark energy models will give different values in terms of the pair $\{S_n,S_n^{(1)}\}$.
The second member of statefinder hierarchy cam be constructed from $S_n^{(1)}$ in the following manner \cite{104}:
\begin{equation}
S^{(2)}_n=\frac{S^{(1)}_n-1}{3(q-\frac{1}{2})} \label{eqs-17}.
\end{equation}
For the $\Lambda$CDM scenario, obviously, $\{S_n,S^{(2)}_n \}=\{1,0 \}$, $\{S^{(1)}_n,S^{(2)}_n \}=\{1,0 \}$. For the dynamical dark energy models, one will get different results so as to distinguish them from the $\Lambda$CDM scenario more conveniently. According to \cite{104}, $\omega$CDM, Chaplygin gas (CG), and DGP model have been discriminated from each other and the base cosmology.
\subsection{The Growth Rate of Matter Perturbations}
The growth rate of perturbations can be acted as an important and effective supplement for the statefinders, and the fractional growth parameter $\epsilon(z)$ is defined as
\begin{equation}
\epsilon(z)=\frac{f(z)}{f_{\Lambda CDM}(z)},
\end{equation}
where
\begin{equation}
f(z)=\Omega_m(z)^{\gamma(z)},
\end{equation}
representing the growth rate of linearized density perturbations. For slowly varying equation of state with time, which satisfy the condition $\left|d\omega/d\Omega_m\right|\ll(1-\Omega_m)^{-1}$, one can get the relationship
\begin{equation}
\gamma(z)=\frac{3}{5-\frac{\omega}{1-\omega}}+\frac{3}{125}\frac{(1-\omega)(1-1.5\omega)}{(1-1.2\omega)^3}[1-\Omega_m(z)]+\mathcal{O}[(1-\Omega_m(z))]^2.
\end{equation}
The above approximation works reasonably well for physical dark energy models with either a constant or a slowly varying equation of state with time. However, it is not the case in modified gravities where the perturbation growth contains information which is just complementary to that contained in the expansion history. For the $\Lambda$CDM scenario, it is easy to find that
\begin{equation}
\epsilon(z)\mid_{\Lambda CDM}=1.
\end{equation}
\begin{figure}
\centering
\includegraphics[scale=0.5]{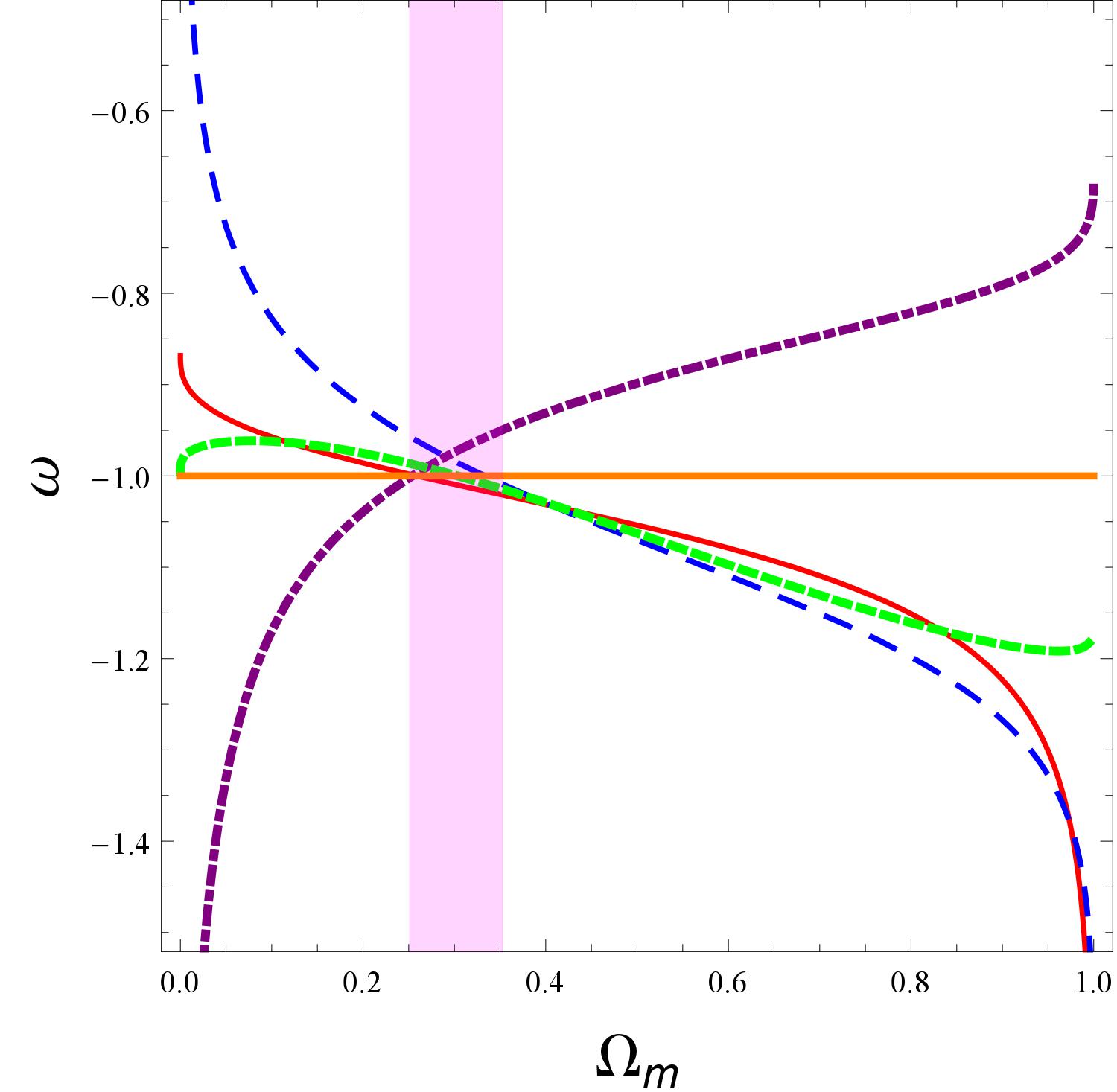}
\caption{The relation between the matter density parameter and the equation of state parameter. The orange (horizontal) line, the red (solid) line, the blue (long-dashed) line, the purple (dash-dotted) line and the green (dashed) line corresponds to the $\Lambda$CDM model, model 1, model 2, model 3 and model 4, respectively. The vertical band centered at $\Omega_{m0}=0.3$ roughly corresponds to the present stage.}\label{1}
\end{figure}
\begin{figure}
\centering
\includegraphics[scale=0.5]{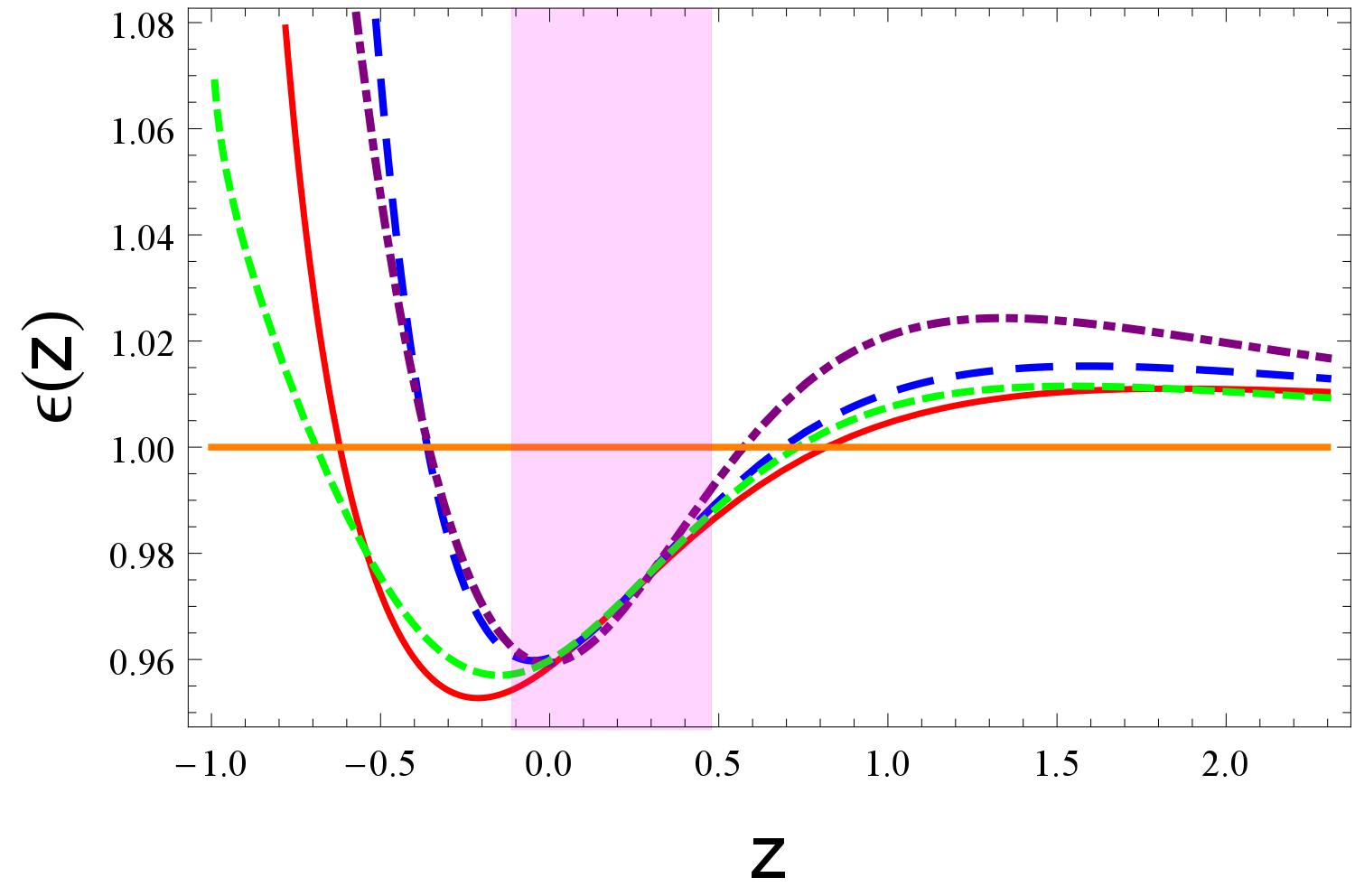}
\caption{The relation between the redshift and the fractional growth parameter $\epsilon(z)$. The orange (horizontal) line, the red (solid) line, the blue (long-dashed) line, the purple (dash-dotted) line and the green (dashed) line corresponds to the $\Lambda$CDM model, model 1, model 2, model 3 and model 4, respectively. The vertical band centered at $z=0$ roughly corresponds to the present stage.}\label{2}
\end{figure}
\begin{figure}
\centering
\includegraphics[scale=0.5]{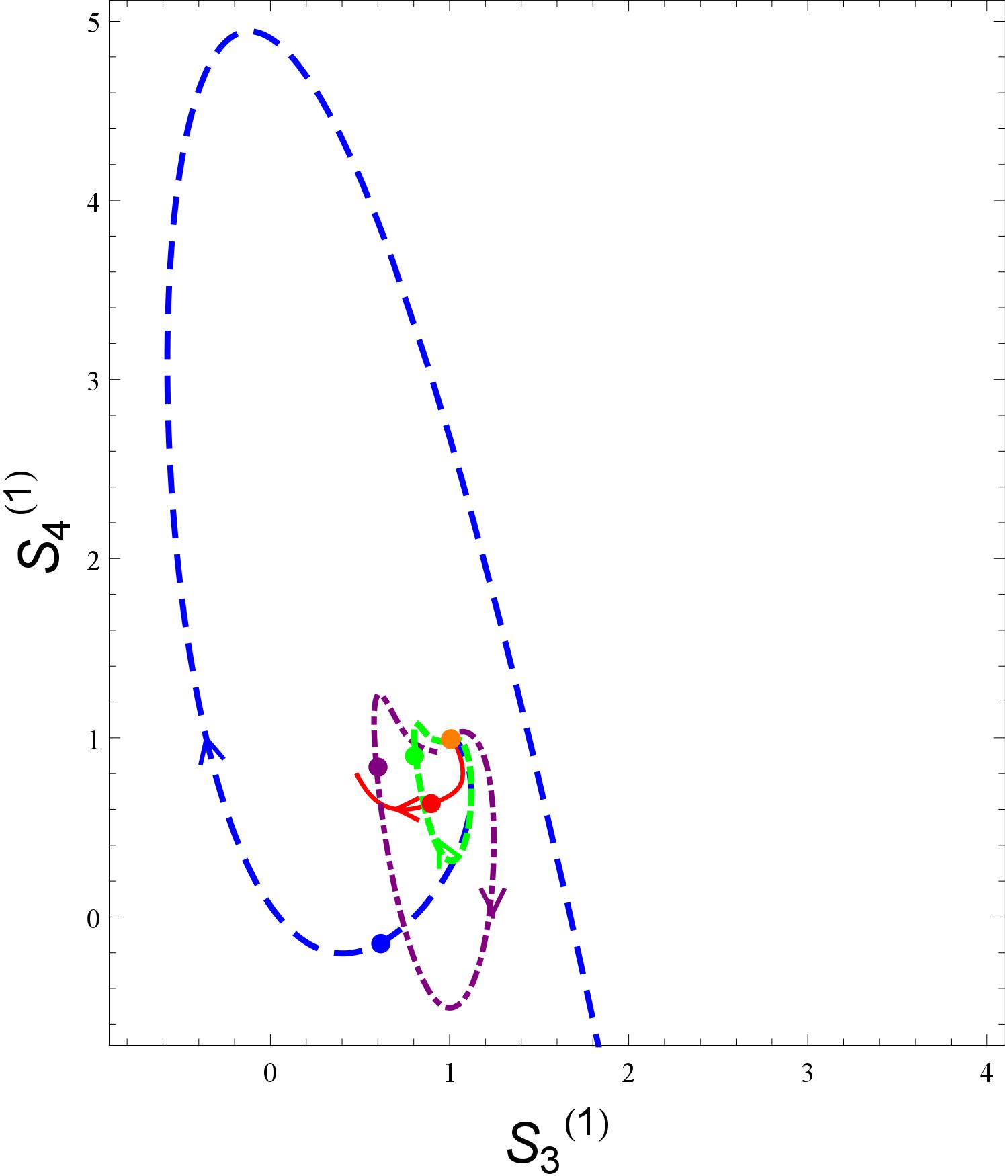}
\caption{The statefinder $\{S_3^{(1)},S_4^{(1)}\}$ plane. The red (solid) line, the blue (long-dashed) line, the purple (dash-dotted) line and the green (dashed) line corresponds to model 1, model 2, model 3 and model 4, respectively. The present epoch in different models is shown as a dot and the arrows imply the evolutional direction with respect to time. The orange dot corresponding to the fixed point $\{1,1\}$ represents the $\Lambda$CDM model.}\label{3}
\end{figure}
\begin{figure}
\centering
\includegraphics[scale=0.52]{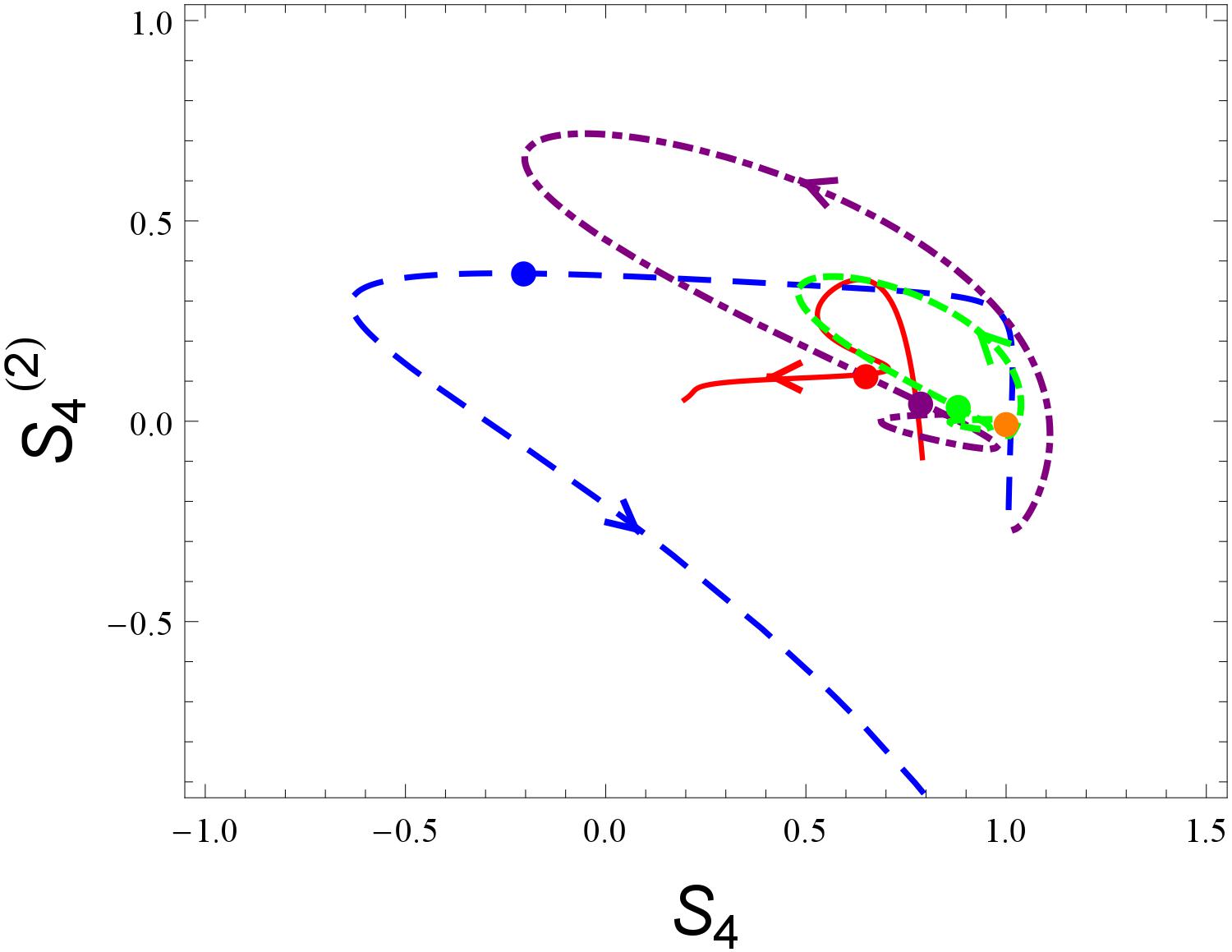}
\caption{The statefinder $\{S_4,S_4^{(2)}\}$ plane. The red (solid) line, the blue (long-dashed) line, the purple (dash-dotted) line and the green (dashed) line corresponds to model 1, model 2, model 3 and model 4, respectively. The present epoch in different models is shown as a dot and the arrows imply the evolutional direction with respect to time. The orange dot corresponding to the fixed point $\{1,0\}$ represents the $\Lambda$CDM model.}\label{4}
\end{figure}
\begin{figure}
\centering
\includegraphics[scale=0.5]{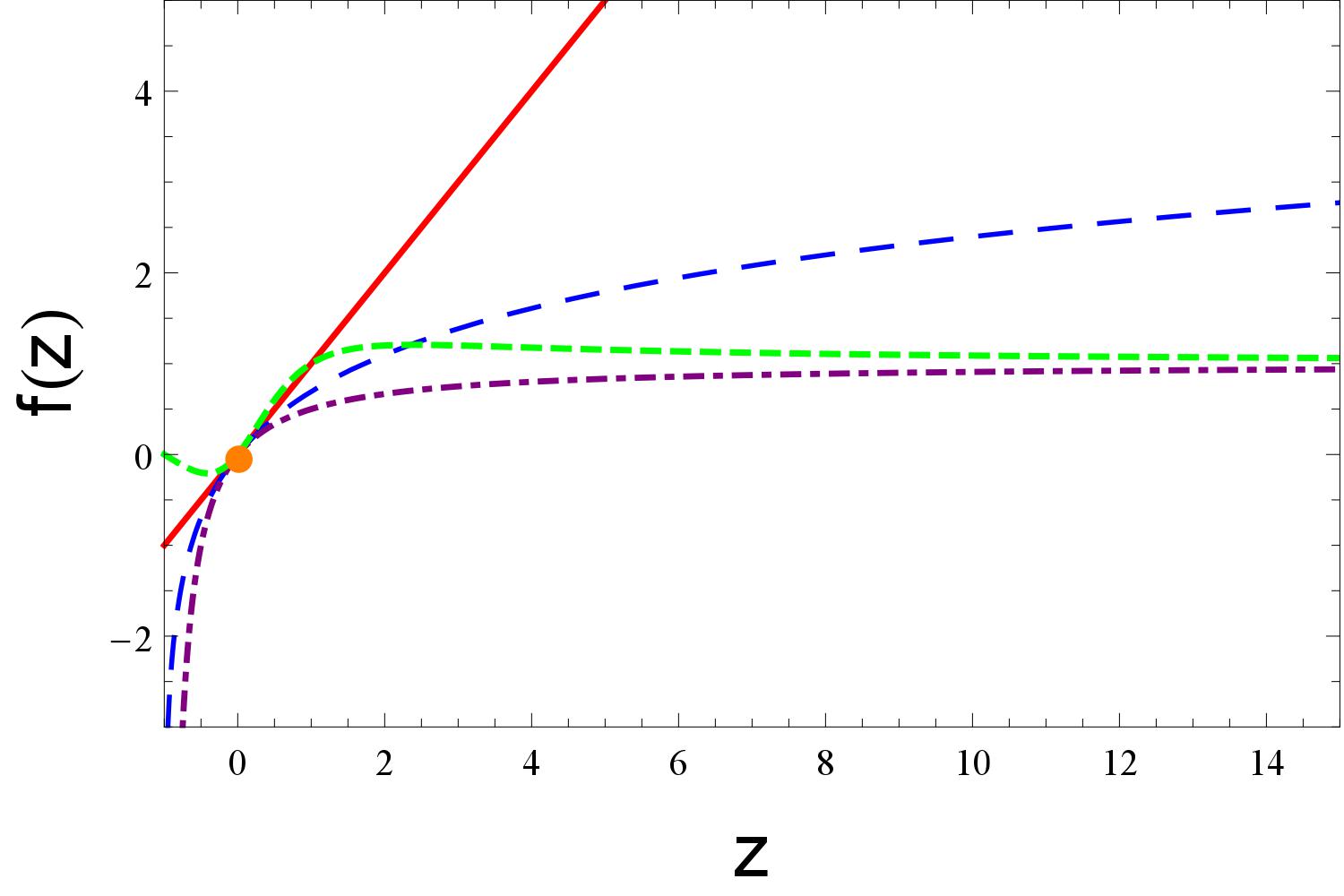}
\caption{The relation between the redshift and the parameterization function. The red (solid) line, the blue (long-dashed) line, the purple (dash-dotted) line and the green (dashed) line corresponds to model 1, model 2, model 3 and model 4, respectively. The $\Lambda$CDM model is shown as a dot which corresponds to the fixed point $\{0,0\}$.}\label{5}
\end{figure}
\begin{figure}
\centering
\includegraphics[scale=0.5]{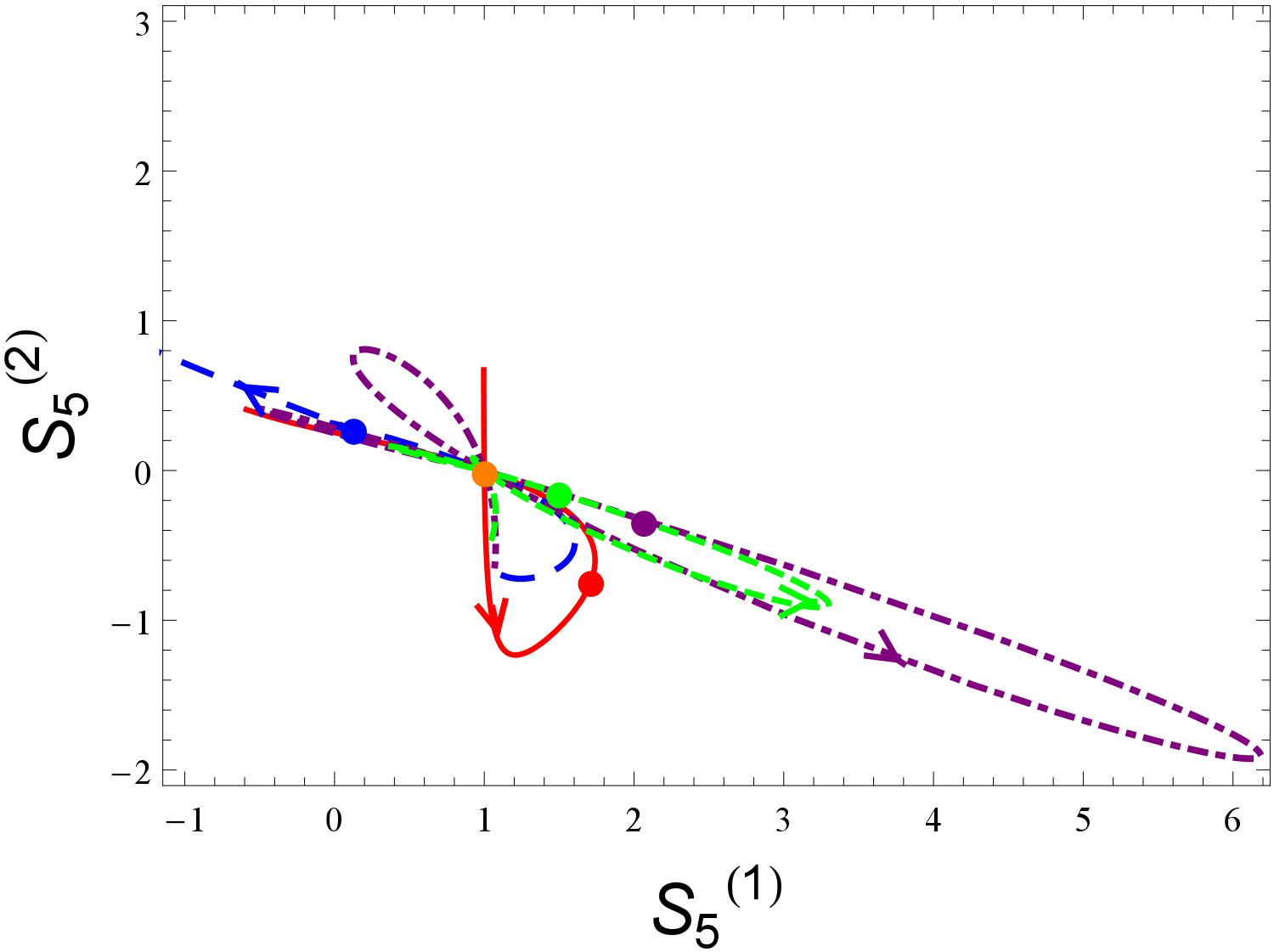}
\caption{The statefinder $\{S_5^{(1)},S_5^{(2)}\}$ plane. The red (solid) line, the blue (long-dashed) line, the purple (dash-dotted) line and the green (dashed) line corresponds to model 1, model 2, model 3 and model 4, respectively. The present epoch in different models is shown as a dot and the arrows imply the evolutional direction with respect to time. The orange dot corresponding to the fixed point $\{1,0\}$ represents the $\Lambda$CDM model.}\label{6}
\end{figure}
\begin{figure}
\centering
\includegraphics[scale=0.5]{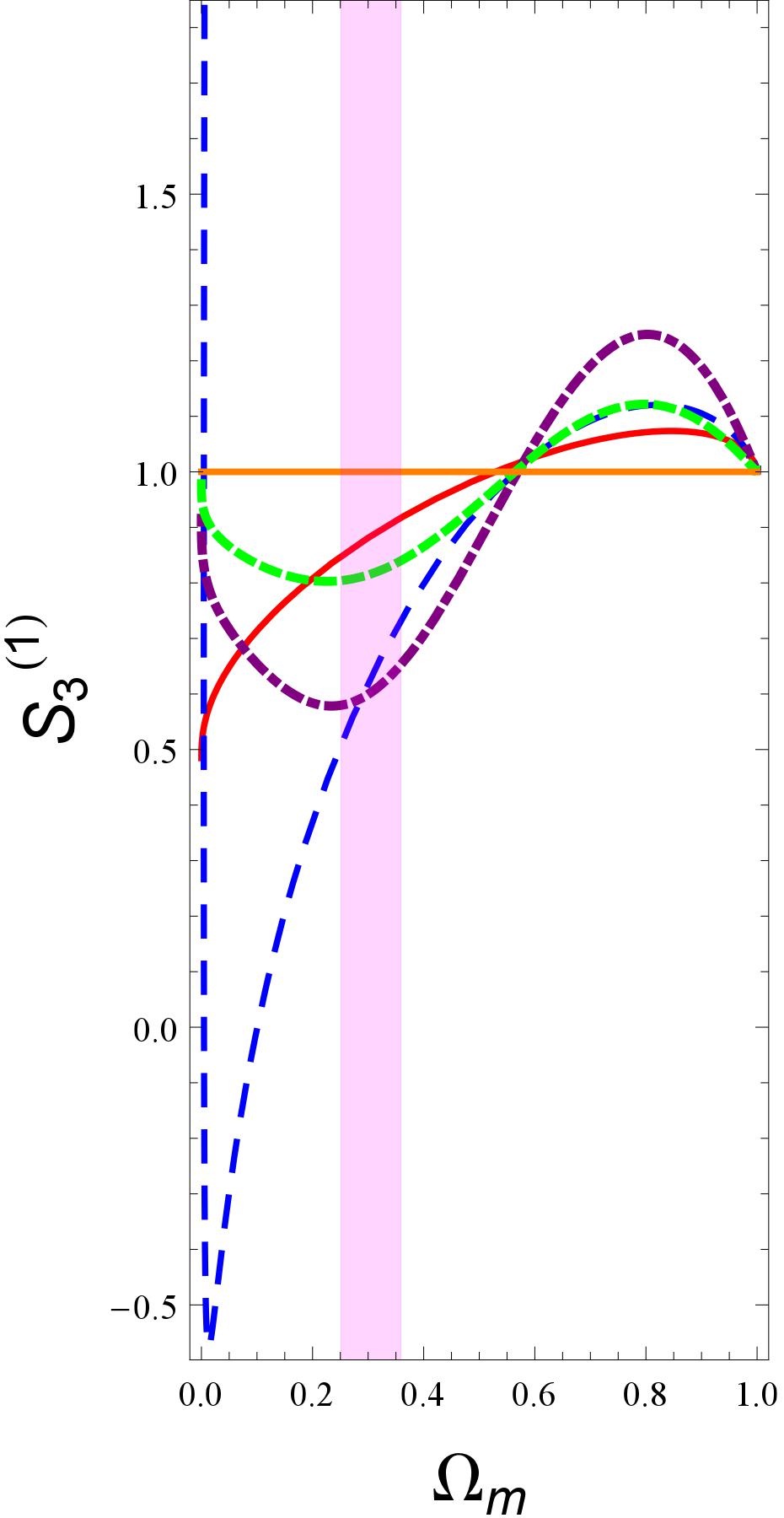}
\caption{The relation between the matter density parameter and the statefinder $S_3^{(1)}$. The orange (horizontal) line, the red (solid) line, the blue (long-dashed) line, the purple (dash-dotted) line and the green (dashed) line corresponds to the $\Lambda$CDM model, model 1, model 2, model 3 and model 4, respectively. The vertical band centered at $\Omega_{m0}=0.3$ roughly corresponds to the present stage.}\label{7}
\end{figure}
\begin{figure}
\centering
\includegraphics[scale=0.5]{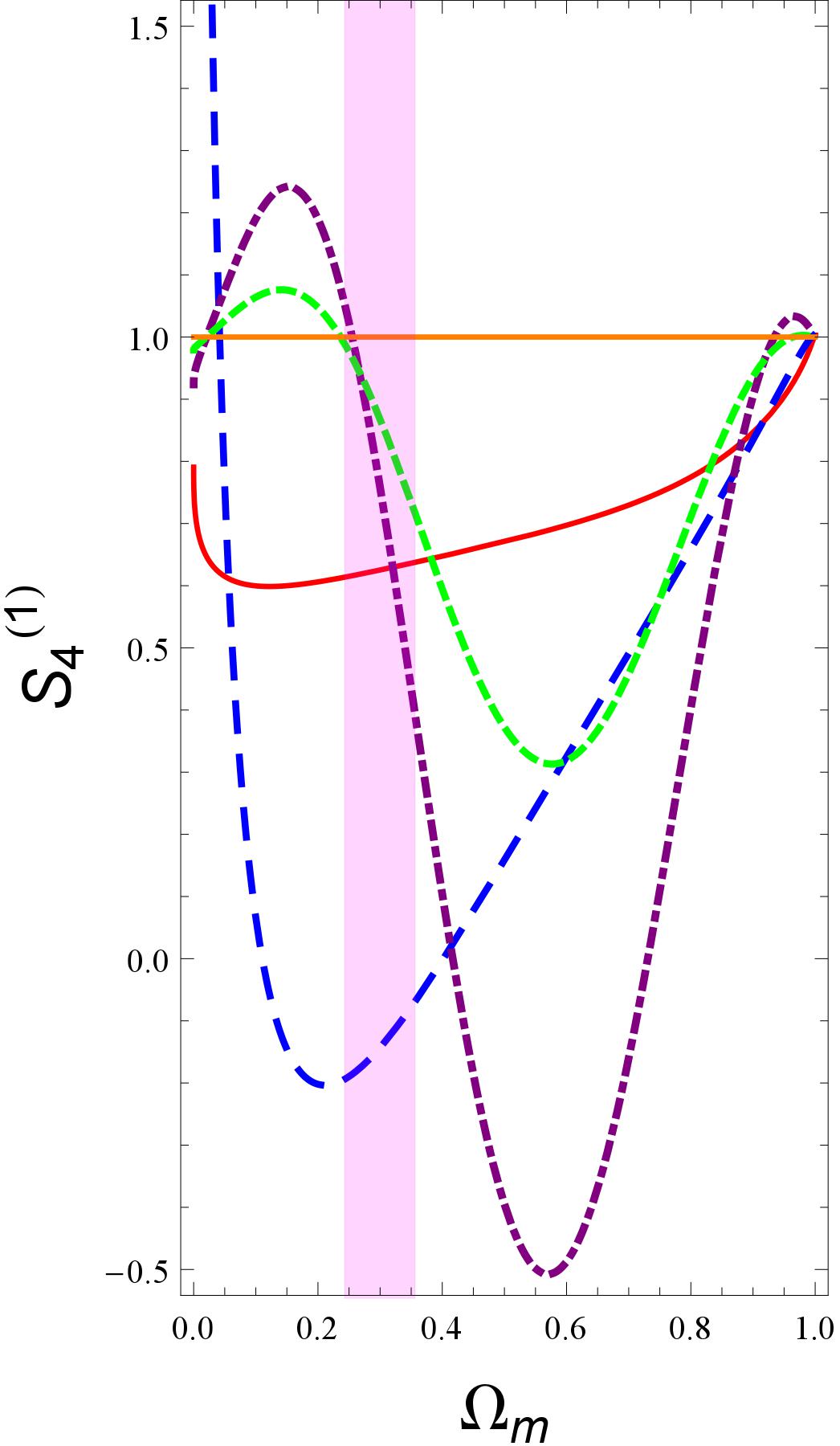}
\caption{The relation between the matter density parameter and the statefinder $S_4^{(1)}$. The orange (horizontal) line, the red (solid) line, the blue (long-dashed) line, the purple (dash-dotted) line and the green (dashed) line corresponds to the $\Lambda$CDM model, model 1, model 2, model 3 and model 4, respectively. The vertical band centered at $\Omega_{m0}=0.3$ roughly corresponds to the present stage.}\label{8}
\end{figure}
\begin{figure}
\centering
\includegraphics[scale=0.5]{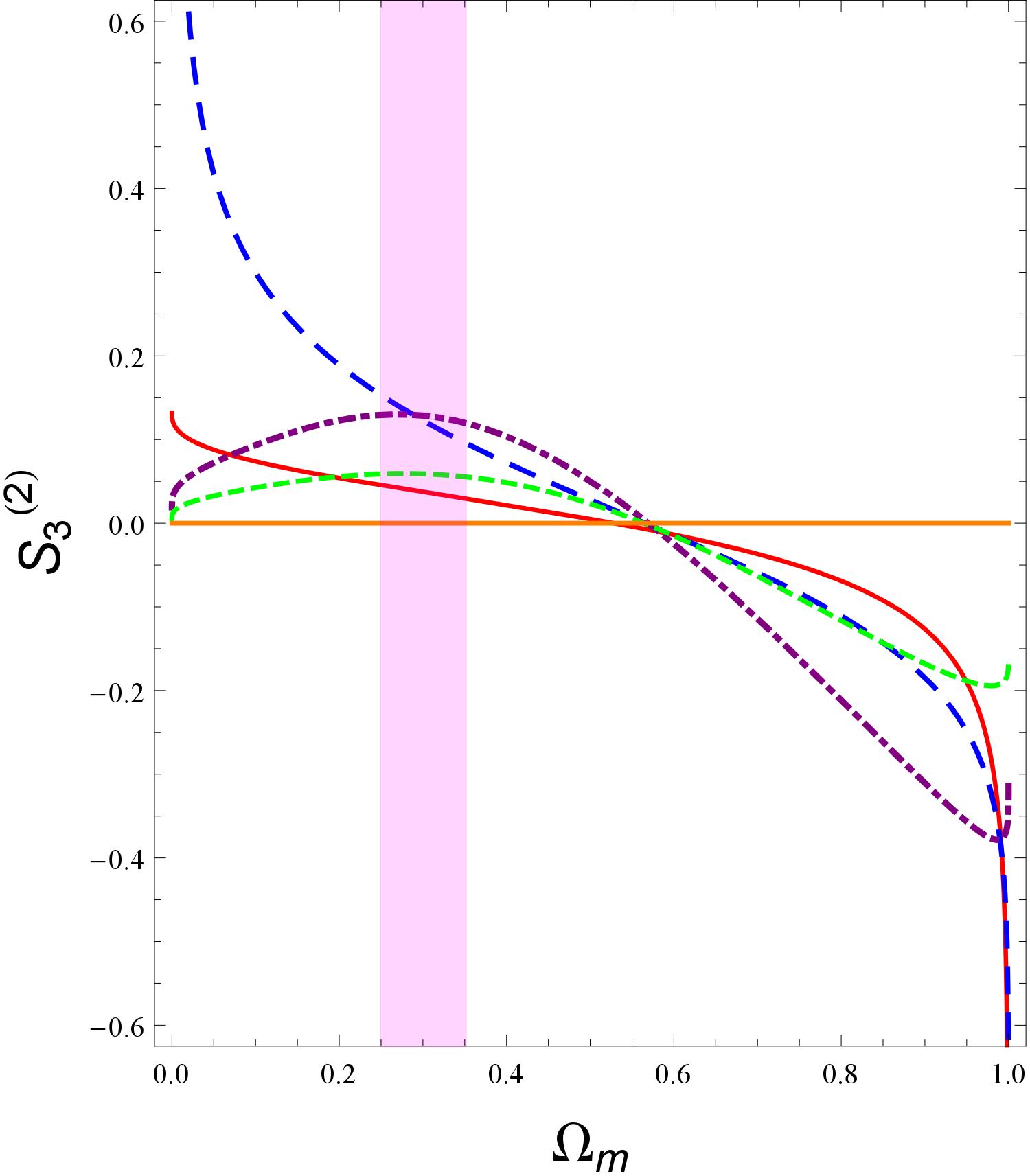}
\caption{The relation between the matter density parameter and the statefinder $S_3^{(2)}$. The orange (horizontal) line, the red (solid) line, the blue (long-dashed) line, the purple (dash-dotted) line and the green (dashed) line corresponds to the $\Lambda$CDM model, model 1, model 2, model 3 and model 4, respectively. The vertical band centered at $\Omega_{m0}=0.3$ roughly corresponds to the present stage.}\label{9}
\end{figure}
\begin{figure}
\centering
\includegraphics[scale=0.5]{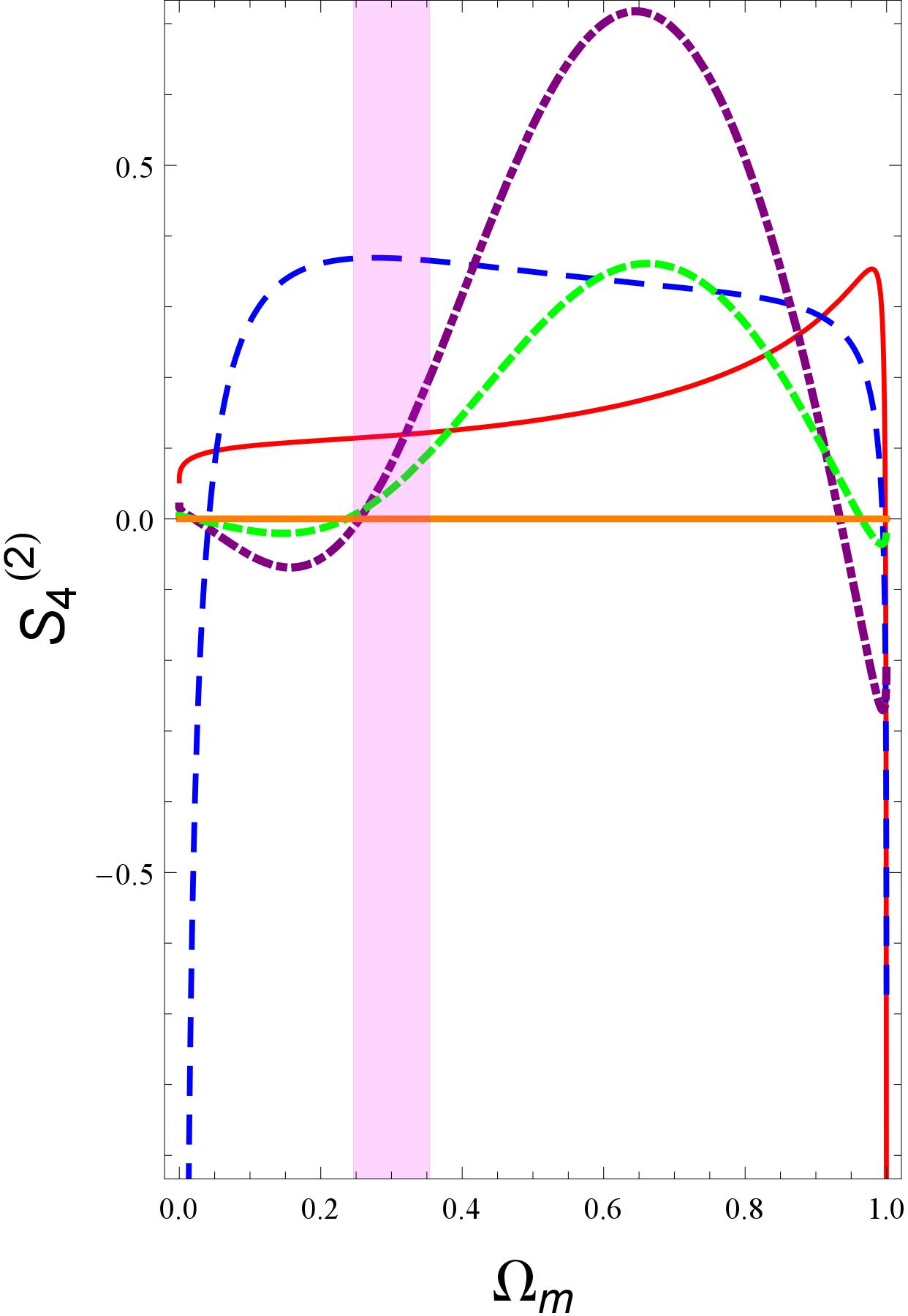}
\caption{The relation between the matter density parameter and the statefinder $S_4^{(2)}$. The orange (horizontal) line, the red (solid) line, the blue (long-dashed) line, the purple (dash-dotted) line and the green (dashed) line corresponds to the $\Lambda$CDM model, model 1, model 2, model 3 and model 4, respectively. The vertical band centered at $\Omega_{m0}=0.3$ roughly corresponds to the present stage.}\label{10}
\end{figure}
\begin{figure}
\centering
\includegraphics[scale=0.5]{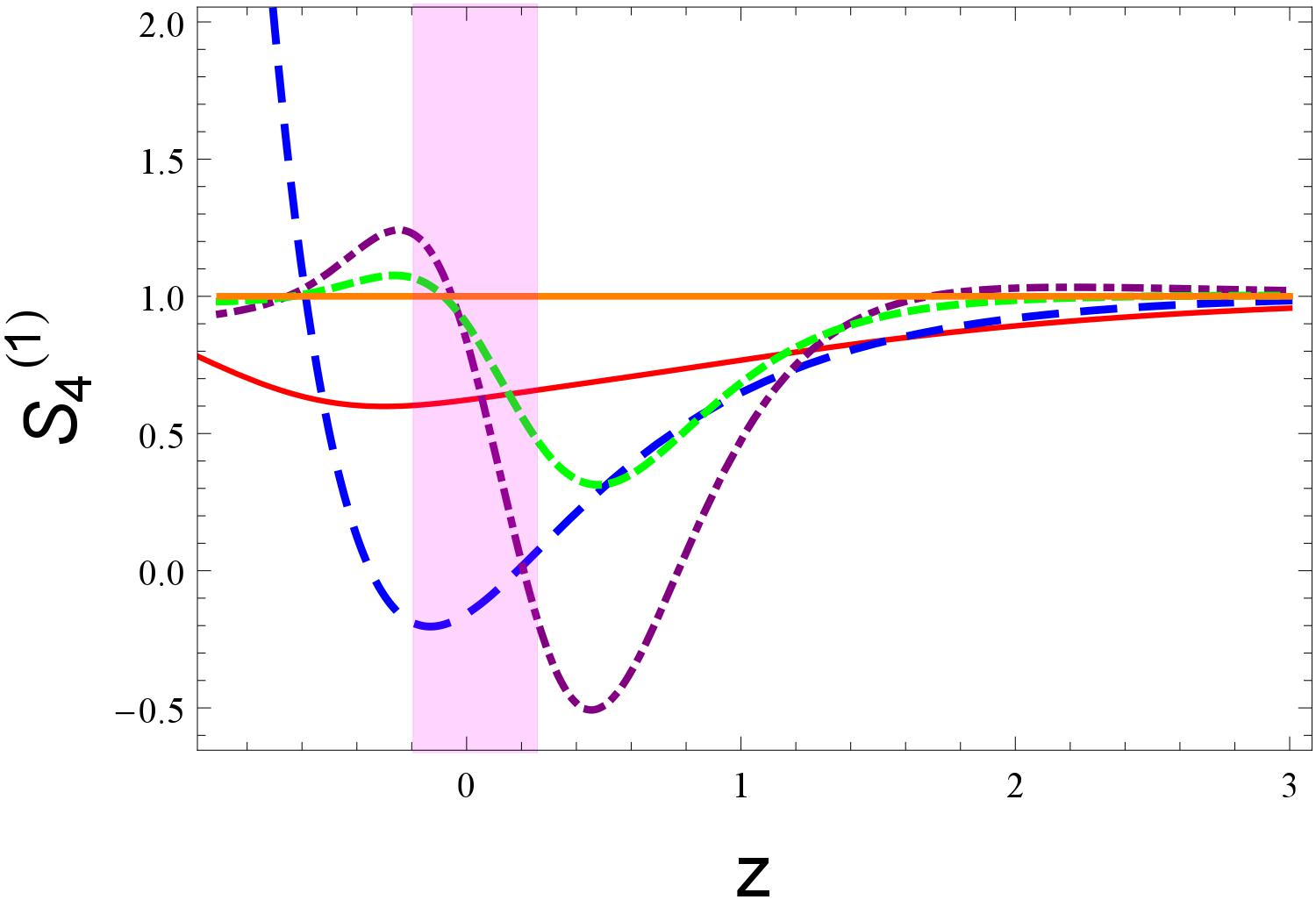}
\caption{The relation between the redshift and the statefinder $S_4^{(1)}$. The orange (horizontal) line, the red (solid) line, the blue (long-dashed) line, the purple (dash-dotted) line and the green (dashed) line corresponds to the $\Lambda$CDM model, model 1, model 2, model 3 and model 4, respectively. The vertical band centered at $z=0$ roughly corresponds to the present stage.}\label{11}
\end{figure}
\begin{figure}
\centering
\includegraphics[scale=0.5]{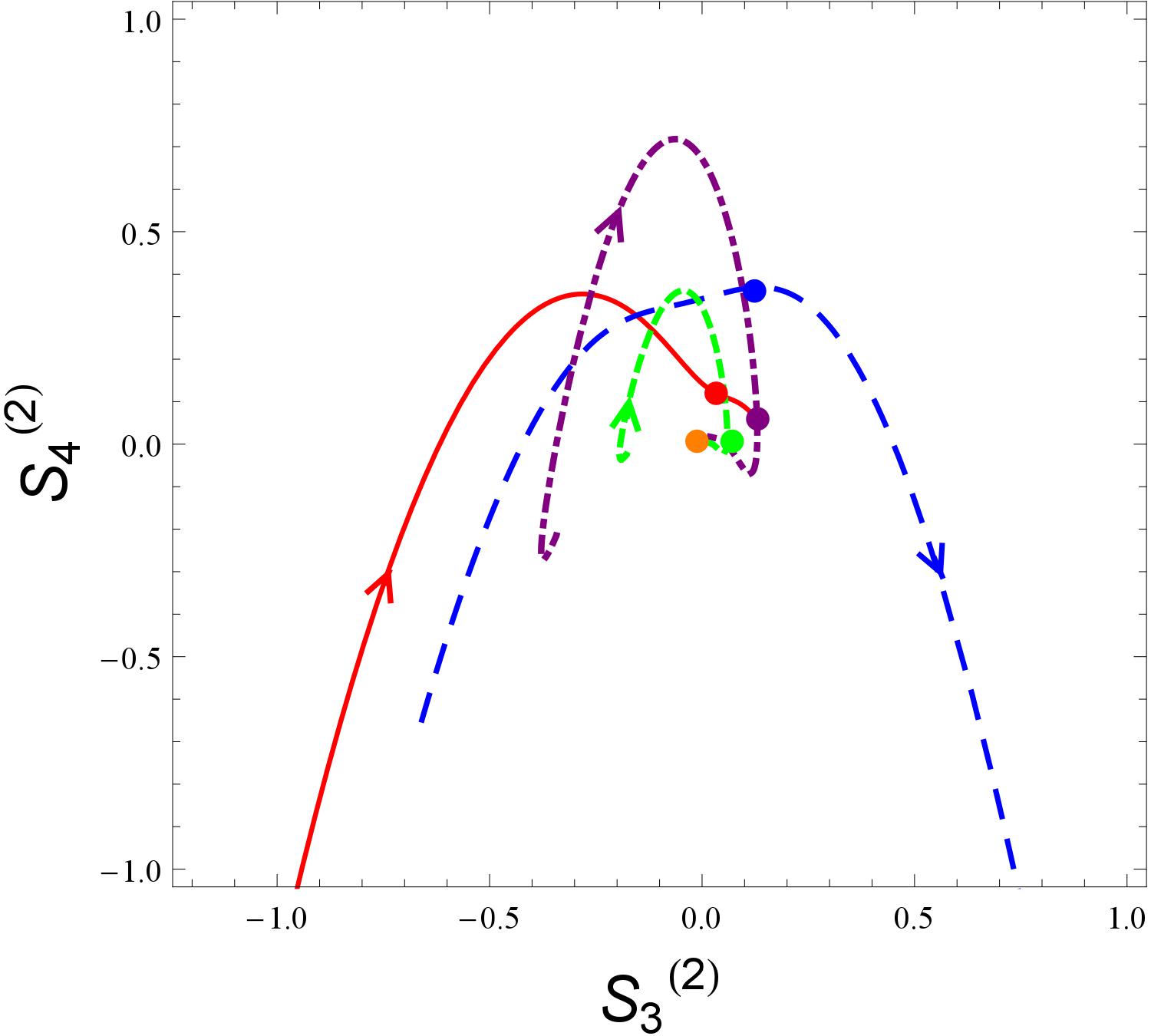}
\caption{The statefinder $\{S_3^{(2)},S_4^{(2)}\}$ plane. The red (solid) line, the blue (long-dashed) line, the purple (dash-dotted) line and the green (dashed) line corresponds to model 1, model 2, model 3 and model 4, respectively. The present epoch in different models is shown as a dot and the arrows imply the evolutional direction with respect to time. The orange dot corresponding to the fixed point $\{0,0\}$ represents the $\Lambda$CDM model.}\label{12}
\end{figure}
\begin{figure}
\centering
\includegraphics[scale=0.5]{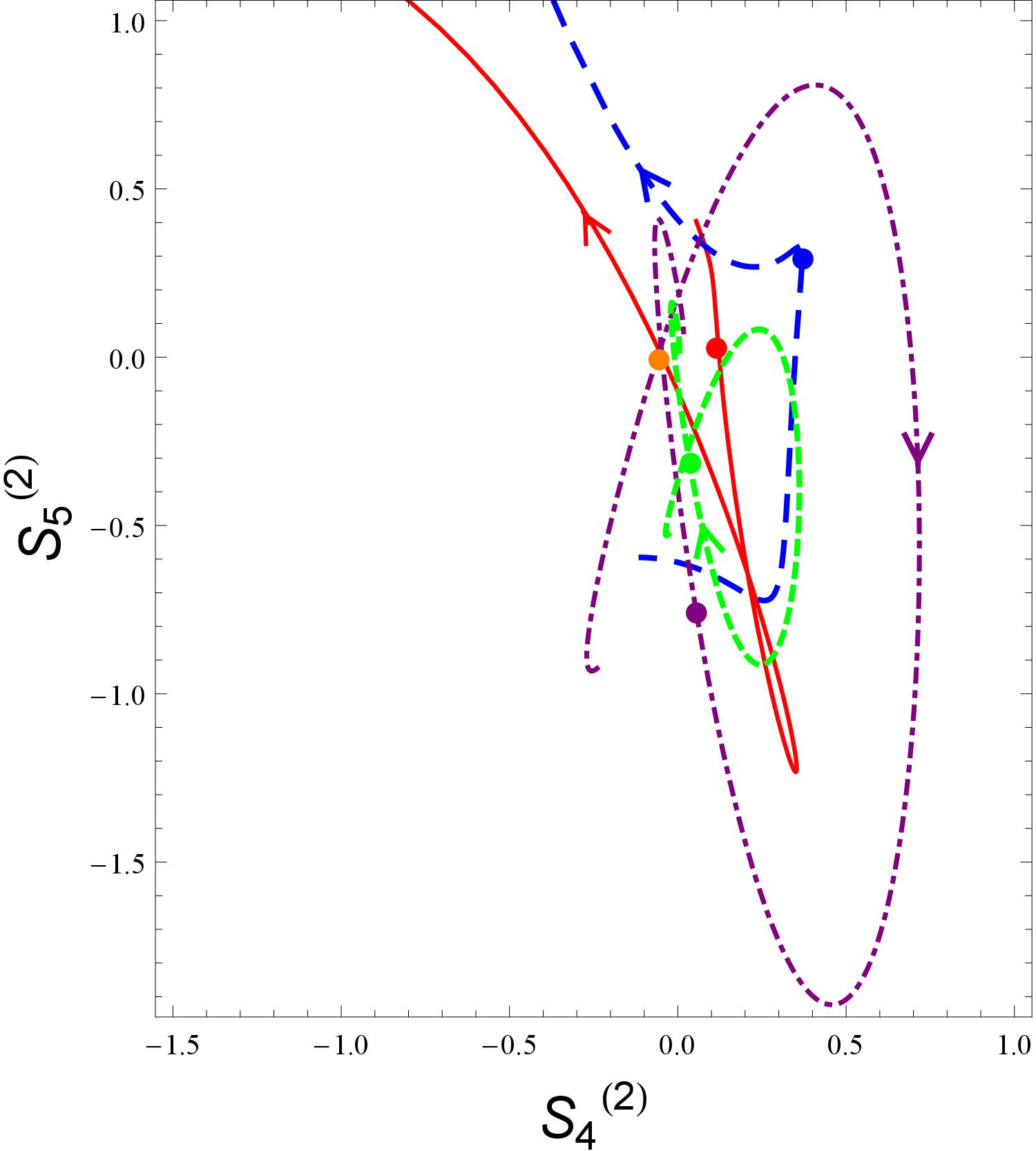}
\caption{The statefinder $\{S_4^{(2)},S_5^{(2)}\}$ plane. The red (solid) line, the blue (long-dashed) line, the purple (dash-dotted) line and the green (dashed) line corresponds to model 1, model 2, model 3 and model 4, respectively. The present epoch in different models is shown as a dot and the arrows imply the evolutional direction with respect to time. The orange dot corresponding to the fixed point $\{0,0\}$ represents the $\Lambda$CDM model.}\label{13}
\end{figure}
\begin{figure}
\centering
\includegraphics[scale=0.5]{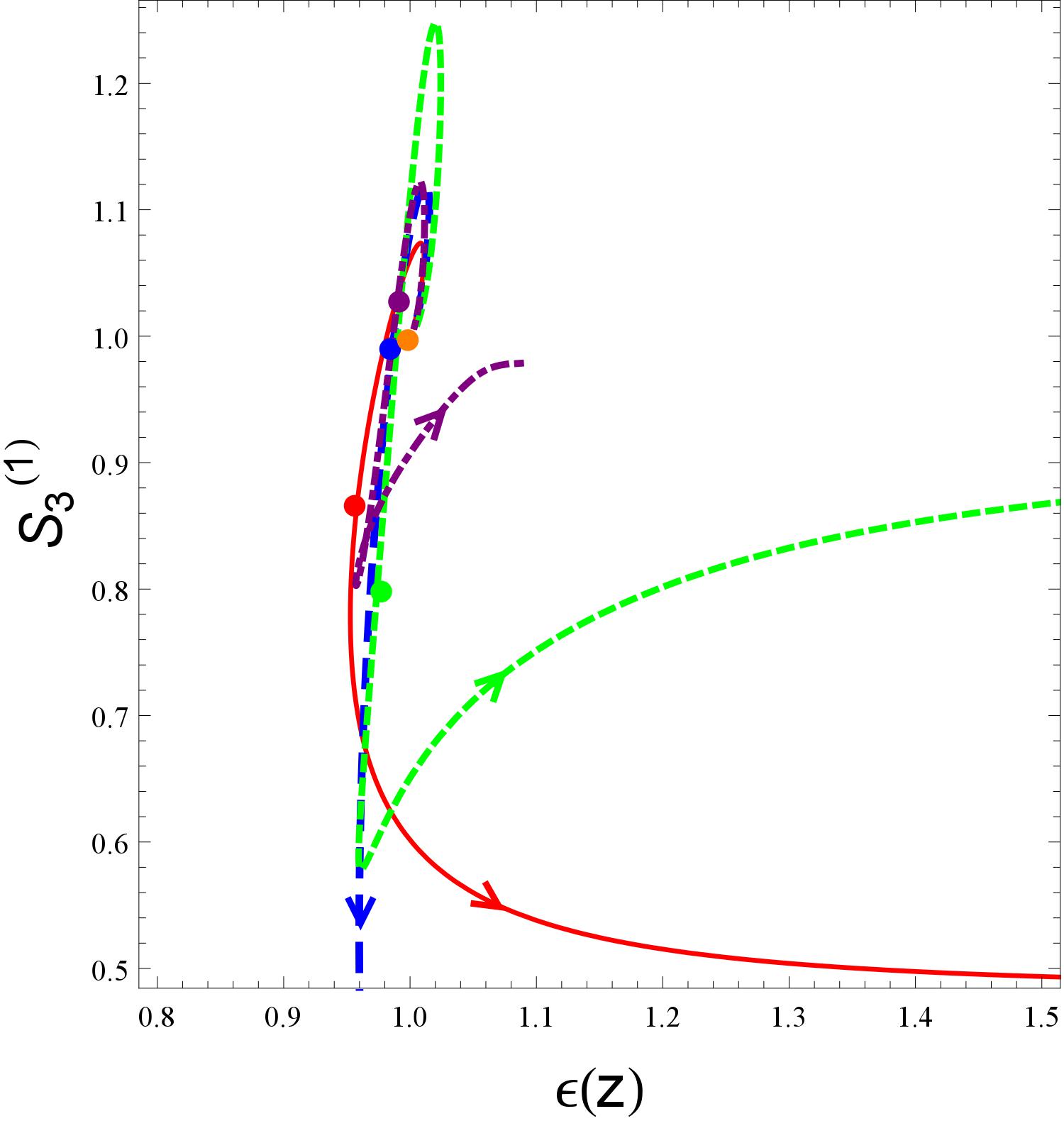}
\caption{The relation between the fractional growth parameter and the statefinder $S_3^{(1)}$. The red (solid) line, the blue (long-dashed) line, the purple (dash-dotted) line and the green (dashed) line corresponds to model 1, model 2, model 3 and model 4, respectively. The present epoch in different models is shown as a dot and the arrows imply the evolutional direction with respect to time. The orange dot corresponding to the fixed point $\{1,1\}$ represents the $\Lambda$CDM model.}\label{14}
\end{figure}
\begin{figure}
\centering
\includegraphics[scale=0.5]{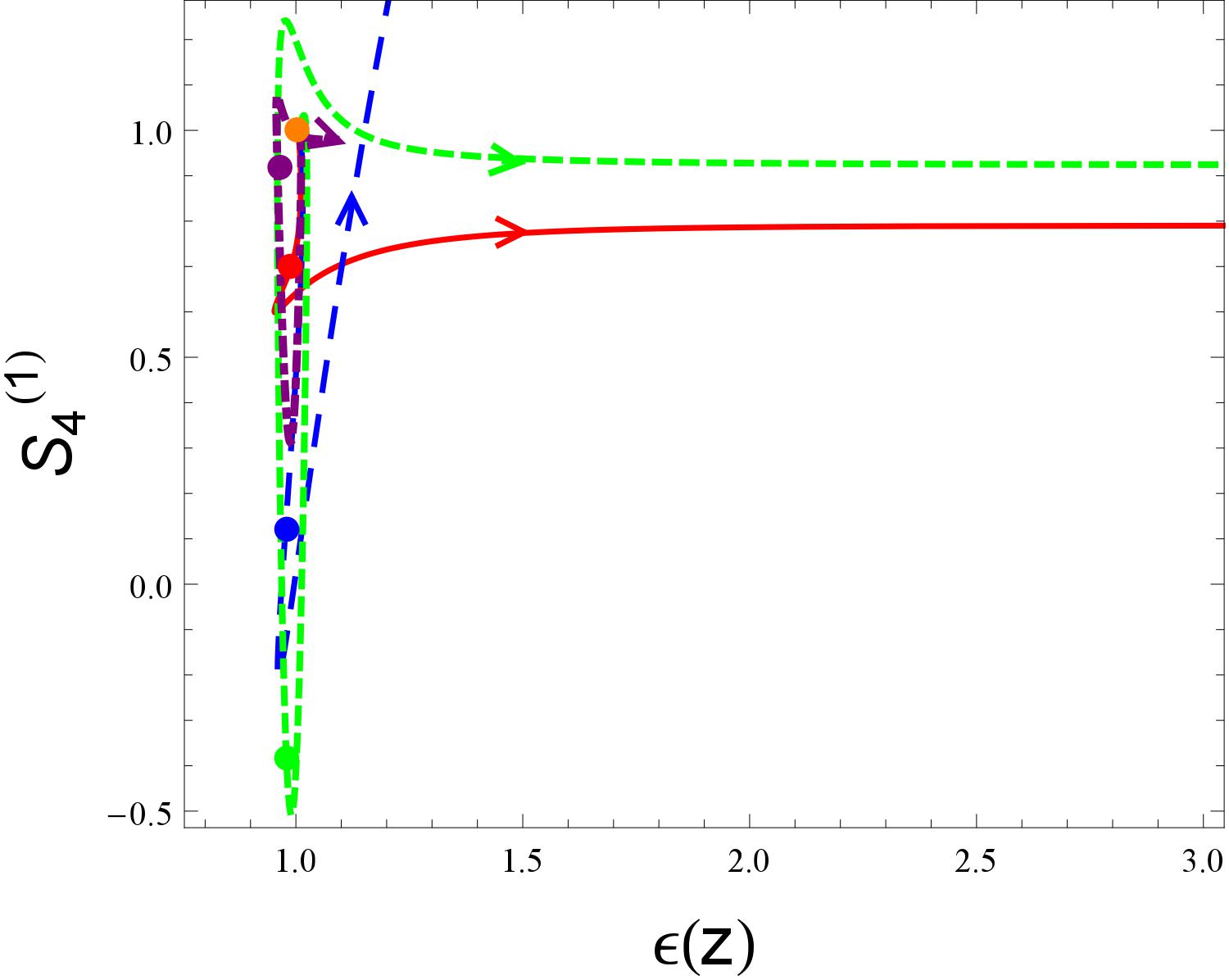}
\caption{The relation between the fractional growth parameter and the statefinder $S_4^{(1)}$. The red (solid) line, the blue (long-dashed) line, the purple (dash-dotted) line and the green (dashed) line corresponds to model 1, model 2, model 3 and model 4, respectively. The present epoch in different models is shown as a dot and the arrows imply the evolutional direction with respect to time. The orange dot corresponding to the fixed point $\{1,1\}$ represents the $\Lambda$CDM model.}\label{15}
\end{figure}
\begin{figure}
\centering
\includegraphics[scale=0.5]{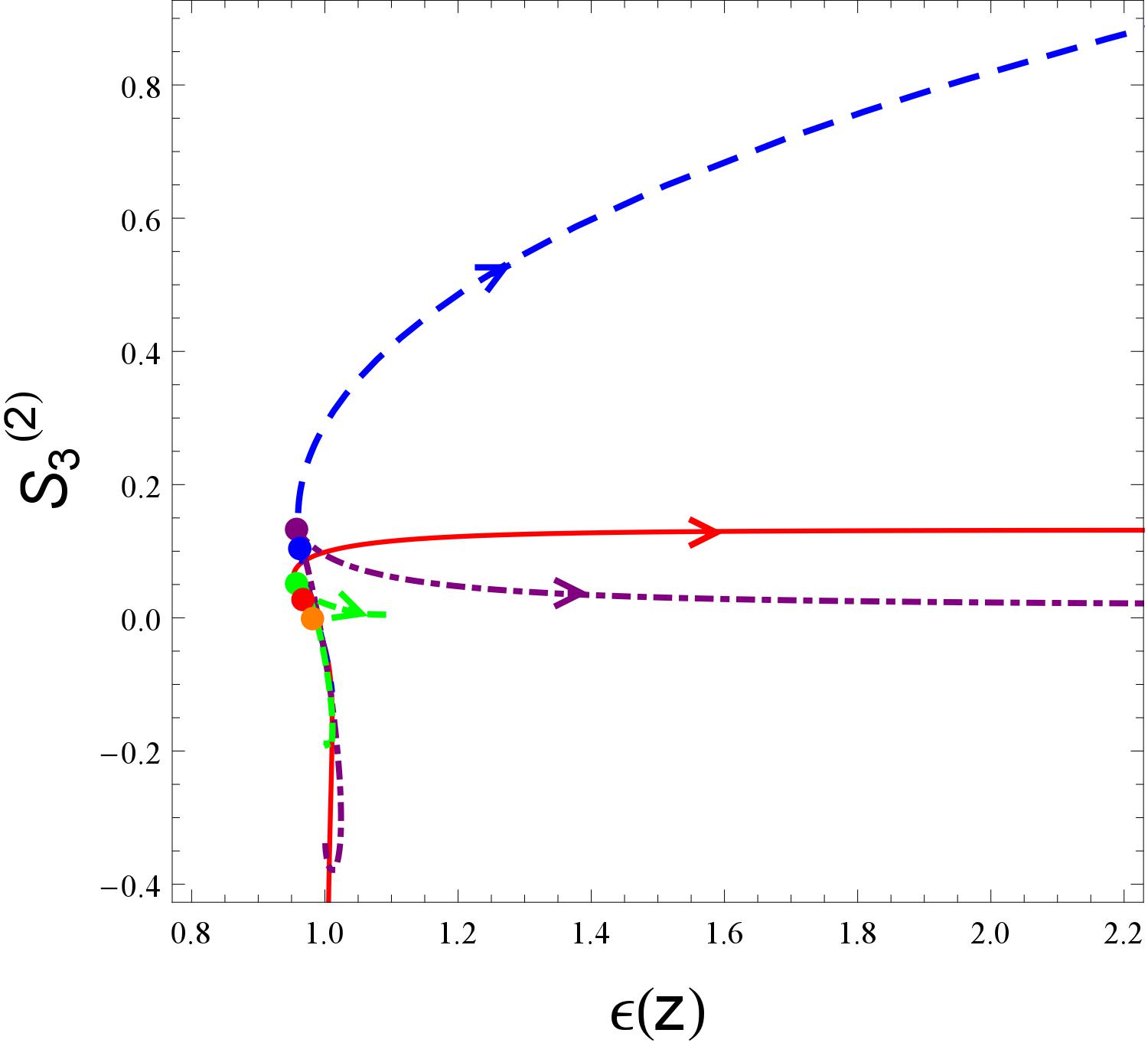}
\caption{The relation between the fractional growth parameter and the statefinder $S_3^{(2)}$. The red (solid) line, the blue (long-dashed) line, the purple (dash-dotted) line and the green (dashed) line corresponds to model 1, model 2, model 3 and model 4, respectively. The present epoch in different models is shown as a dot and the arrows imply the evolutional direction with respect to time. The orange dot corresponding to the fixed point $\{1,0\}$ represents the $\Lambda$CDM model.}\label{16}
\end{figure}
\begin{figure}
\centering
\includegraphics[scale=0.5]{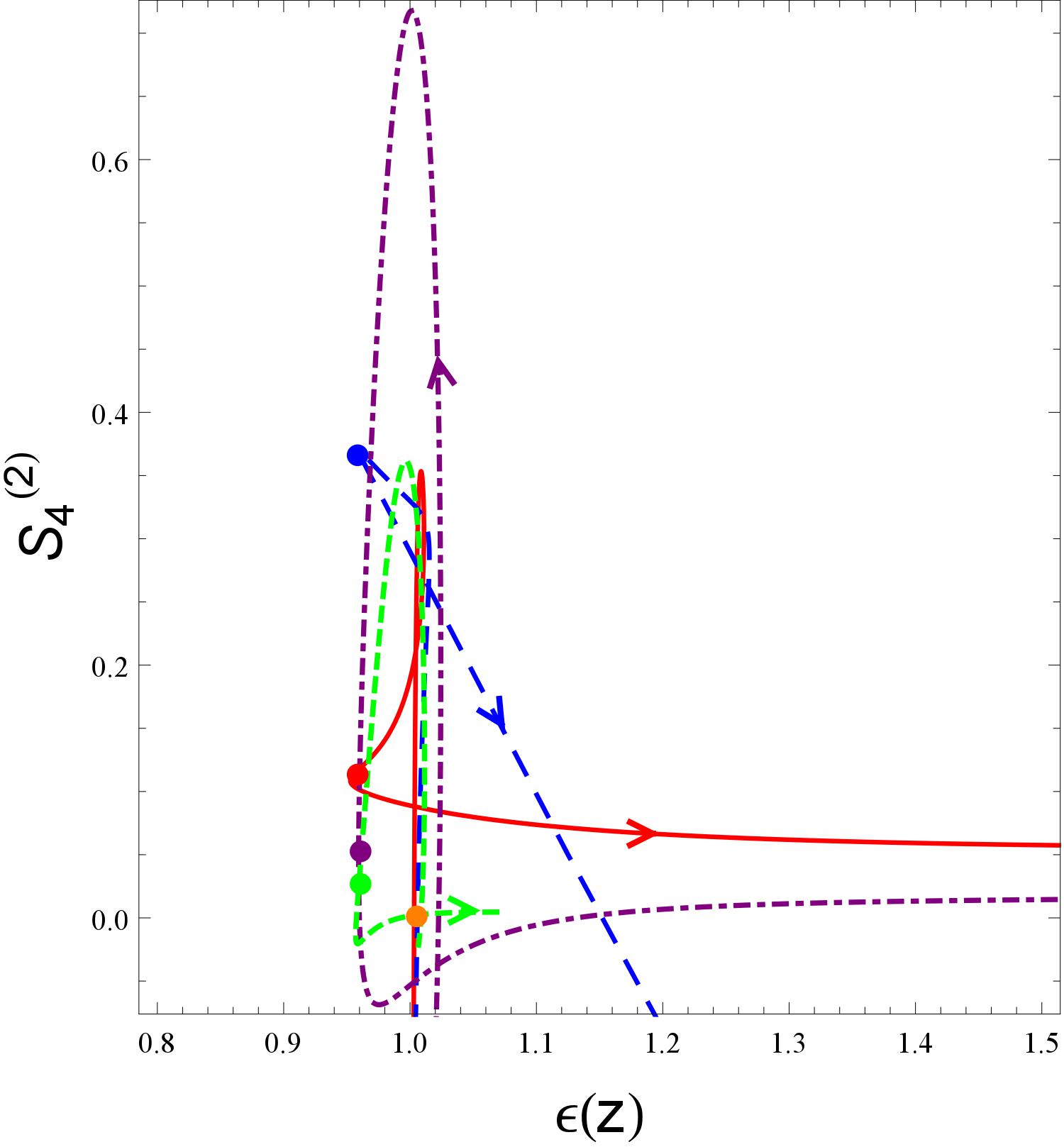}
\caption{The relation between the fractional growth parameter and the statefinder $S_4^{(2)}$. The red (solid) line, the blue (long-dashed) line, the purple (dash-dotted) line and the green (dashed) line corresponds to model 1, model 2, model 3 and model 4, respectively. The present epoch in different models is shown as a dot and the arrows imply the evolutional direction with respect to time. The orange dot corresponding to the fixed point $\{1,0\}$ represents the $\Lambda$CDM model.}\label{17}
\end{figure}
\begin{figure}
\centering
\includegraphics[scale=0.5]{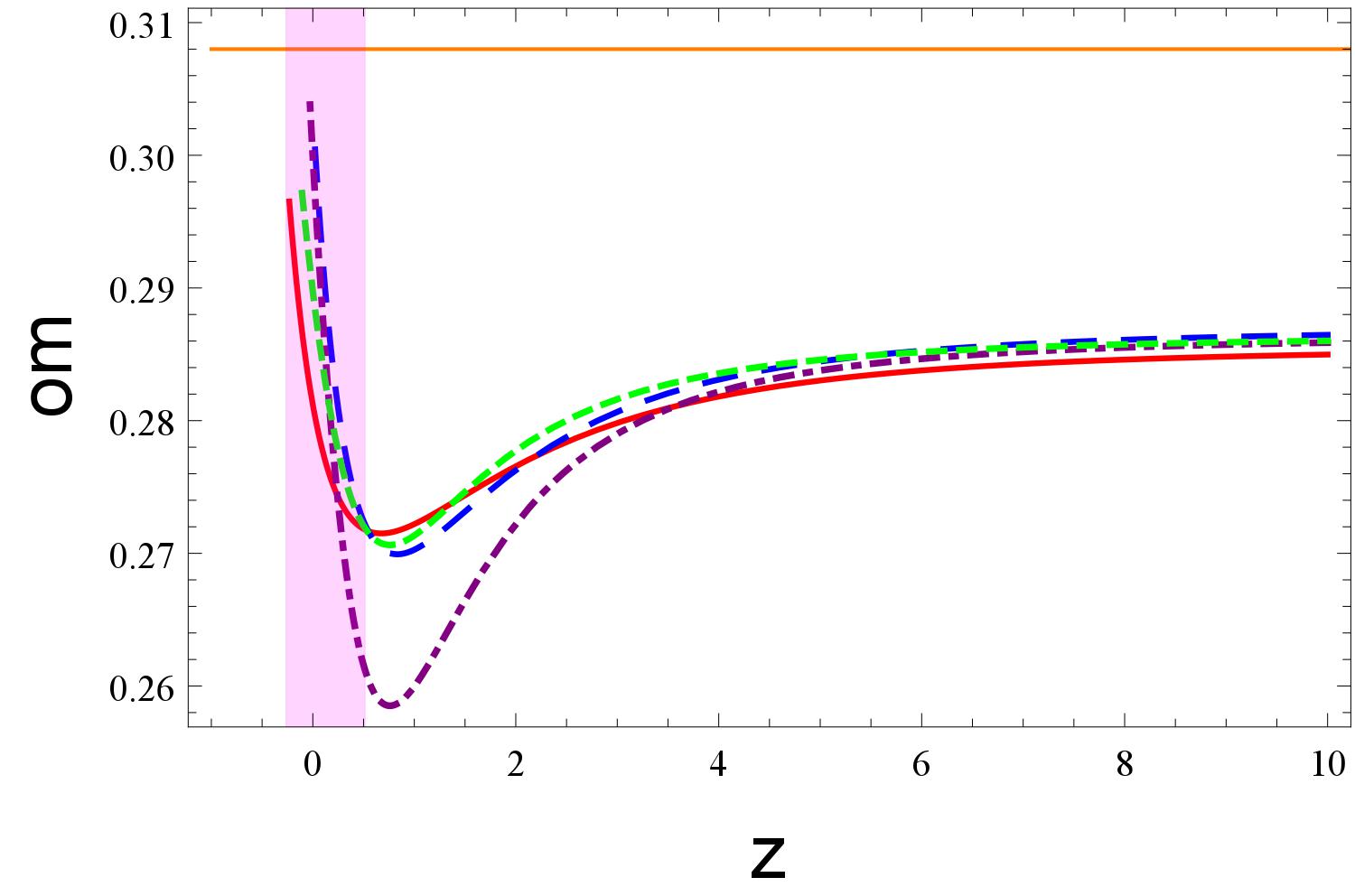}
\caption{The relation between the redshift and $Om(z)$. The orange (horizontal) line, the red (solid) line, the blue (long-dashed) line, the purple (dash-dotted) line and the green (dashed) line corresponds to the $\Lambda$CDM model, model 1, model 2, model 3 and model 4, respectively. The vertical band centered at $z=0$ roughly corresponds to the present stage.}\label{18}
\end{figure}
Thus, one can combine the statefinders and the the growth rate of perturbations to define a composite null diagnostic (CND): $\{\epsilon(z),S_n\}$, $\{\epsilon(z),S_n^{(1)}\}$ or $\{\epsilon(z),S_n^{(2)}\}$. By adopting $\{\epsilon(z),S_3^{(1)}\}$, $\omega$CDM, DGP model and the $\Lambda$CDM model have been  well distinguished in \cite{104}. By adopting $\{\epsilon(z),S_4\}$, $\{\epsilon(z),S_3^{(1)}\}$ and $\{\epsilon(z),S_5^{(1)}\}$, GCG, MCG, SCG, PKK and the $\Lambda$CDM model have already been well distinguished in \cite{93}.
\subsection{The $Om(z)$ Diagnostic}
The $Om(z)$ diagnostic is also an useful method to discriminate various dark energy models, and can be defined as
\begin{equation}
Om(x)=\frac{E^2(x)-1}{x^3-1} \label{eqs-23},
\end{equation}
where $E(x)=H(x)/H_0$ and $x=1/a=1+z$. Similarly, neglecting the radiation at low redshifts, for the base cosmology, one can easily obtain
\begin{equation}
E^2(x)=\Omega_{m0}x^3+(1-\Omega_{m0})  \label{eqs-24}.
\end{equation}
Substituting Eq. (\ref{eqs-24}) into Eq. (\ref{eqs-23}), one can get
\begin{equation}
Om(x)\mid_{\Lambda CDM}=\Omega_{m0} \label{eqs-25}.
\end{equation}
It is not difficult to find that the $Om(z)$ diagnostic also provides a null test for the base cosmology, and for other evolving dark energy models, the $Om(z)$ diagnostics are expected to give different values. In our previous work \cite{77}, the two parametrization models for effective pressure have been well distinguished from each other and the $\Lambda$CDM model.
\section{Discriminations with the statefinder hierarchy, the growth rate of matter perturbations and the $Om(z)$ diagnostic}
In the following context, we would like to apply the statefinder hierarchy, the growth rate of matter perturbations and the $Om(z)$ diagnostic into discriminating the aforementioned dark energy models. According to \cite{104}, the parameters $q$, $A_3$, $A_4$ and $A_5$ can be expressed as
\begin{eqnarray}
  q&=&(1+z)\frac{1}{E}\frac{dE}{dz}-1 \label{eqs-19}, \\
  A_3&=&(1+z)\frac{1}{E^2}\frac{d[E^2(1+q)]}{dz}-3q-2 \label{eqs-20},   \\
  A_4&=&-(1+z)\frac{1}{E^3}\frac{d[E^3(2+3q+A_3)]}{dz}+4A_3+3q(q+4)+6 \label{eqs-21},   \\
  A_5&=&-(1+z)\frac{1}{E^4}\frac{d[E^4(A_4-4A_3-3q(q+4)-6)]}{dz}+5A_4-10A_3(q+2)-30q(q+2)-24.  \label{eqs-22}
\end{eqnarray}

As mentioned above, Arabsalmani et al. \cite{104} have used $\{S_3^{(1)},S_4^{(1)}\}$ and $\{\epsilon(z),S_3^{(1)}\}$, respectively, to discriminate the CG, $\omega$CDM, DPG and the $\Lambda$CDM model, and $\omega$CDM, DPG, and the $\Lambda$CDM model. In addition, Li et al. have already used the statefinder $\{S_3^{(1)},S_4^{(1)}\}$, $\{S_4,S_4^{(1)}\}$ and $\{S_3^{(1)},S_5\}$, and the CND $\{\epsilon(z),S_3^{(1)}\}$, $\{\epsilon(z),S_4\}$ and $\{\epsilon(z),S_5^{(1)}\}$ to discriminate GCG, MCG, SCG, PKK and the $\Lambda$CDM model. In this situation, we also adopt the statefinder $\{S_3^{(1)},S_4^{(1)}\}$ to distinguish the four TDDE models and the $\Lambda$CDM scenario from each other. In Figure. \ref{3}, we have plotted the evolutional trajectories of the aforementioned TDDE models in the plane of $\{S_3^{(1)},S_4^{(1)}\}$. It is easy to be seen that all the models can be well distinguished from each other at the present stage. In particular, the trajectory of model 2 is completely different from the left three models. Additionally, one can discover that the evolutional tendency of the models 3 and 4 is very similar.

Before applying the the growth rate of matter perturbations to discriminate the four TDDE models, through some numerical calculations, we discover that all the mentioned-above TDDE models satisfy the slowly varying condition at the present stage. To be more precise, when $z=0$, for model 1, we find $\left|d\omega/d\Omega_m\right|=0.34274\ll(1-\Omega_m)^{-1}\approx1.39957$, $\left|d\omega/d\Omega_m\right|=0.24591\ll(1-\Omega_m)^{-1}\approx1.40243$ for model 2, $\left|d\omega/d\Omega_m\right|=0.41763\ll(1-\Omega_m)^{-1}\approx1.40125$ for model 3, and $\left|d\omega/d\Omega_m\right|=0.08595\ll(1-\Omega_m)^{-1}\approx1.40128$ for model 4. Moreover, one can easily find that the values of $(1-\Omega_m)^{-1}$ for these four models are very close to each other.

In Fig. \ref{1}, we have plotted the evolutional trajectories in the $\Omega_m-\omega$ plane for the four TDDE models and the $\Lambda$CDM model. Obviously, for models 1, 2 and 4, there exists a high degeneracy in the substantially long period ($0.2\lesssim\Omega_m\lesssim0.9$) and vary very slowly at the present epoch. For model 3, one can find that the trajectory corresponds to a monotonically decreasing function $\omega(\Omega_m)$, and vary more slowly at the present stage than in the remote past.

In Figure. \ref{2}, we have plotted the evolutional behavior of the fractional growth parameter $\epsilon(z)$ for the aforementioned TDDE scenarios and the $\Lambda$CDM scenario. It is easy to be seen that these four TDDE models just run closely to the $\Lambda$CDM model in the remote past, and deviate obviously from the base cosmology when $z<1.5$. Furthermore, one can find that these models obey a high degeneracy so as to be hardly distinguished at the present epoch.

Using the statefinder $\{S_4,S_4^{(2)}\}$, it is obvious that all the models will deviate gradually from the base cosmology with time, and also model 2 can be discriminated better from the base cosmology than the other models at the present epoch. At the same time, models 3 and 4 still exhibit the similar evolution tendency. Hence, it is worth investigating the function formalism $f(z)$ for the four TDDE parameterization models very much. From Figure. \ref{5}, one can discover that the parameterizations for the equation of state of models 3 and 4 are very similar in the substantially long period, which can provide an excellent and reasonable explanation for the similar evolution tendency of the two models in the planes of $\{S_3^{(1)},S_4^{(1)}\}$ and $\{S_4,S_4^{(2)}\}$.

In the plane of statefinder $\{S_5^{(1)},S_5^{(2)}\}$, one can also find that these four models can be well distinguished from each other and the $\Lambda$CDM model at the present epoch. More importantly, one can get the following interesting conclusion: models 1, 3 and 4 will go through the fixed point $\{1,0\}$ which corresponds to the $\Lambda$CDM scenario more than one time. Subsequently, we plot the evolutional trajectories of these models in the planes of $\{\Omega_m,S_3^{(1)}\}$, $\{\Omega_m,S_4^{(1)}\}$, $\{\Omega_m,S_3^{(2)}\}$ and $\{\Omega_m,S_4^{(2)}\}$ in order to understand the attractive phenomenon better. It is not difficult to find that the evolutional trajectories of these TDDE models in these figures go through the horizontal line which corresponds to    the $\Lambda$CDM scenario more than one time. Furthermore, by plotting the vertical band centered at $\Omega_{m0}=0.3$ which corresponds roughly to the present stage, one can find that these models may not be well distinguished from each other and the $\Lambda$CDM scenario. For instance, in Figure. \ref{7}, models 2 and 3 can not be well discriminated at the present epoch since the two evolutional trajectories share one overlap, and in Figure. \ref{8}, one can also observe two overlaps, which means models 3 and 4 can not be distinguished at the present stage, so do models 1 and 3. In Figure. \ref{9}, one still distinguish models 2 and 3 as in Figure. \ref{7} at the present stage. In Figure. \ref{10}, model 3 remains not discriminated from models 1 and 4 at the present epoch, but can be well distinguished form the $\Lambda$CDM scenario. Then, we think that it is constructive to exhibit the evolutional behavior of the statefinder hierarchy. As a concrete example, we plot the evolutional trajectory of the TDDE models in the $\{z,S_4^{(1)}\}$ plane (see Figure. \ref{11}).

In Figures. \ref{12} and \ref{13}, we adopt the statefinder $\{S_3^{(2)},S_4^{(2)}\}$ and $\{S_4^{(2)},S_5^{(2)}\}$ to discriminate the TDDE models. From Figure. \ref{12}, it is easy to be seen that model 2 can be well distinguished from the left three models and the $\Lambda$CDM model at the present epoch, and model 4 may not be well distinguished from the $\Lambda$CDM model. Moreover, models 3 and 4 will approach the $\Lambda$CDM scenario in the far future. From Figure. \ref{13}, one will discover that all the TDDE models can be distinguished at the present stage better than in Figure. \ref{12}, and models 1 and 3 will be the same with the base cosmology, respectively, at two different epochs during the evolution of the universe.

Subsequently, we will use the CND to distinguish the four TDDE models and the $\Lambda$CDM model. In Figure. \ref{14}, one can easily find that all the TDDE models evolve starting from the $\Lambda$CDM model and gradually deviate from each other and the base cosmology. In addition, one may discover that model 2 can be hardly distinguished from the base cosmology at the present epoch. In the $\{\epsilon(z),S_4^{(1)}\}$ plane (see Figure. \ref{15}), one can find that model 3 may not be well discriminated from the $\Lambda$CDM scenario at the present epoch. More appealingly, the distance in terms of $S_4^{(1)}$ between model 1 and model 4 will be invariable in the far future. In Figure. \ref{16}, it is obvious that all the models are hardly distinguished from the $\Lambda$CDM scenario and one from other at the present epoch. Moreover, the evolutional trajectory of model 2 is completely different from other models in the far future. As for Figure. \ref{17}, we find that all the TDDE models can be well discriminated from the $\Lambda$CDM model at the present epoch, but models 3 and 4 may not be well distinguished from each other at the present stage. Additionally, one could find that there exists an apparent break point for model 2 during the evolution of the universe, and the distance in terms of $S_4^{(2)}$ between model 1 and model 3 will tend to be invariable in the remote future.

In Figure. \ref{18}, we adopt the $Om(z)$ diagnostic to distinguish the TDDE models from the $\Lambda$CDM model, and one from other. Obviously, one can discover that the TDDE models can not be distinguished from each other at the present epoch. Actually, comparing with the astrophysical observations, one can not discriminate the TDDE models and $\Lambda$CDM model at the present epoch in terms of $68.3\%$ confidence level.

It is worth noting that we also adopt other statefinder pairs, such as $\{S_3,S_3^{(2)}\}$ and $\{S_3,S_5^{(2)}\}$, to discriminate the TDDE models from each other and the base cosmology, and discover that all these models can not be well distinguished form each other at the present epoch, comparing with the mentioned-above results.

\section{Concluding Remarks}
Since cosmologists have proposed various kinds of dark energy models to explain the accelerated mechanism of the recent universe expansion, it is worth investigating the relationship among different cosmological models. Sahni et al. \cite{91} have constructed a series of diagnostics to distinguish different dark energy models from each other and the base cosmology scenario, including the original statefinder, statefinder hierarchy, CND, $Om(z)$ and $Om3(z)$ diagnostics.

In this paper, first of all, we place constraints on the four (TDDE) models by using the SNe Ia, BAO, OHD data-sets as well as the single data point from the newest event GW150914. Subsequently, we have adopted the former four diagnostics to discriminate four TDDE models from each other and the base cosmology scenario. As mentioned above, we have plotted the evolutional trajectories of these models in the planes of $\{S_3^{(1)},S_4^{(1)}\}$, $\{S_4,S_4^{(2)}\}$, $\{S_5^{(1)},S_5^{(2)}\}$, $\{S_3^{(2)},S_4^{(2)}\}$, $\{S_4^{(2)},S_5^{(2)}\}$, $\{\epsilon(z),S_3^{(1)}\}$, $\{\epsilon(z),S_4^{(2)}\}$, $\{Om,z\}$, etc. Through further detailed analysis, we discover that these four TDDE models and the $\Lambda$CDM model can be well distinguished from each other at the present epoch in the planes of $\{S_3^{(1)},S_4^{(1)}\}$, $\{S_3^{(2)},S_4^{(2)}\}$, $\{S_5^{(1)},S_5^{(2)}\}$ and $\{S_4^{(2)},S_5^{(2)}\}$. Since models 3 and 4 share a similar evolution tendency in the planes of $\{S_3^{(1)},S_4^{(1)}\}$, $\{S_4,S_4^{(2)}\}$ and $\{S_5^{(1)},S_5^{(2)}\}$, we have plotted the function formalism $f(z)$ of the two parameterizations and discover that the parameterizations for the equation of state of models 3 and 4 are very similar in the substantially long period, which can also provide an excellent and reasonable explanation for the similar evolution tendency of the two models. Furthermore, to understand the phenomenon that models 1, 3 and 4 go through the fixed point $\{1,0\}$ which corresponds to the $\Lambda$CDM scenario more than one time in the statefinder $\{S_5^{(1)},S_5^{(2)}\}$ plane better, we also plot the evolutional trajectories of the TDDE models and the $\Lambda$CDM scenario in the planes of $\{S_3^{(1)},\Omega_m\}$, $\{S_4^{(1)},\Omega_m\}$, $\{S_3^{(2)},\Omega_m\}$, $\{S_4^{(2)},\Omega_m\}$ and $\{S_4^{(1)},z\}$. It is not difficult to be seen that in the aforementioned figures, the evolutional trajectories of these TDDE models go through the horizontal line which corresponds to the $\Lambda$CDM scenario more than one time. Subsequently, using the CND $\{S_3^{(1)},\epsilon(z)\}$, we find that all the TDDE models evolve starting from the $\Lambda$CDM model and gradually deviate from each other, and model 2 can be hardly distinguished from the base cosmology at the present epoch. In the CND $\{S_4^{(1)},\epsilon(z)\}$ plane, we discover that model 3 may not be well distinguished from the $\Lambda$CDM scenario at the present epoch. More attractively, the `` distance '' in terms of $S_4^{(1)}$ between model 1 and model 4 will be almost invariable in the remote future. In the $\{S_4^{(2)},\epsilon(z)\}$ plane, we find an interesting phenomenon, namely, there exists an apparent break point for model 2 during the evolution of the universe. Moreover, in the CND $\{S_3^{(2)},\epsilon(z)\}$ plane, we obtain the conclusion that all the models are hardly distinguished from the $\Lambda$CDM scenario and one from the other at the present epoch, and the evolutional trajectory of model 2 is completely different from other models in the far future. As a supplement, using the $Om(z)$ diagnostic, we discover that the TDDE models can be hardly distinguished from each other at the present epoch. To be more precise, one can not distinguish the TDDE models and $\Lambda$CDM model from each other at the present epoch in terms of $68.3\%$ confidence interval.

Obviously, the outcomes obtained by adopting the statefinder hierarchy can distinguish the TDDE models and $\Lambda$CDM model from each other better than that obtained by adopting the CND. In particular, we discover that it is not the case, namely, the higher order statedinders we adopt, the better one can distinguish various kinds of dark energy models from each other.

In particular, it must be highlighted that the gravitational wave is an elegant and powerful window for new physics. We would like to make the best use of it to constrain various kinds of modified gravity models and dynamical dark energy models.

Our forthcoming work could be to construct new diagnostics which can be also regarded as more effective tool to distinguish various dark energy scenarios from each other. In addition, it is substantially constructive to use the more accurate observations to constrain the TDDE models and other cosmological models in the future.

\section{acknowledgements}
During the preparation process of the present study, we are grateful to Professors Bharat Ratra and Saibal Ray for very interesting communications on gravitational waves physics in cosmology and compact star formation as well as wormhole astrophysics. The author Deng Wang thanks Prof. Jing-Ling Chen for some helpful discussions and Guang Yang for programming.
This study is supported in part by the National Science Foundation of China.


\begin{thebibliography}{999}
\bibitem{1}
Adam~G.~Riess~et~al, Astrophys. J. {\bf 560}, 49 (2001);

\bibitem{2}
S. Perlmutter, M. S. Turner, and M. White, Phys. Rev. Lett. {\bf 83}, 670 (1999);

\bibitem{3}
D. J. Eisenstein et al, Astrophys. J. {\bf 633}, 560 (2005).

\bibitem{4}
S. Weinberg, Rev. Mod. Phy. {\bf 61}, 1 (1989).

\bibitem{5}
Planck Collaboration, [arXiv:1502.01589].

\bibitem{6}
R. R. Caldwell, Phys. Lett. B {\bf 545} 23-29 (2002).

\bibitem{7}
Y. Fujii, Phys. Rev. D {\bf26,} 2580 (1982).

\bibitem{8}
L. H. Ford, Phys. Rev. D {\bf35,} 2339 (1987).

\bibitem{9}
C. Wetterich, Nucl. Phys. B {\bf302,} 668 (1988).

\bibitem{10}
B. Ratra, P. J. E. Peebles, Phys. Rev. D {\bf37,} 3406 (1988).

\bibitem{11}
S. M. Carroll, Phys. Rev. Lett. {\bf81,} 3067 (1998).

\bibitem{12}
A. Hebecker, C. Witterich, Phys. Rev. Lett. {\bf86,} 3339 (2000).

\bibitem{13}
A. Hebecker, C. Witterich, Phys. Lett. B {\bf497,} 281 (2001).

\bibitem{14}
M. S. Turner, [arXiv: astro-ph/0108103].

\bibitem{15}
R. R. Caldwell, M. Kamionkovski and N.N. Weinberg, Phys. Rev. Lett. {\bf 91}, 071301 (2003).

\bibitem{16}
Yi-Fu Cai, Physics Reports {\bf 493}, 1-60 (2010).

\bibitem{23}
Xin-He Meng, J.Ren and M.G.Hu, Commun. Theor. Phys. {\bf47,} 379 (2007), [arXiv:astro-ph/0509250].

\bibitem{24}
J. Ren and Xin-He Meng, Phys. Lett. B {\bf636,} 5 (2006).

\bibitem{25}
J.Ren and Xin-He Meng, Phys. Lett. B {\bf633,} 1 (2006).

\bibitem{26}
M. G. Hu and Xin-He Meng, Phys. Lett. B {\bf635,} 186 (2006).

\bibitem{27}
Xin-He Meng, X. Dou, Commun. Theor. Phys. {\bf52,} 377 (2009).

\bibitem{28}
Xu Dou, Xin-He Meng, Adv. Astron. {\bf1155,} 829340 (2011).

\bibitem{28.1}
Iver H. Brevik et al, Phys. Rev. D {\bf70}, 043520 (2004).

\bibitem{29}
M. Malekjani, Astrophys. Space. Sci. {\bf 334}, 193-201 (2011).

\bibitem{30}
Xianghua Zhai, Int. J. Mod. Phys. D {\bf 8}, 1151-1161 (2006).

\bibitem{31}
B. PourHassan, Int. J. Mod. Phys. D {\bf 9}, 1350061 (2013).

\bibitem{32}
Jianbo Lu et al, JHEP {\bf 2}, 71 (2015).

\bibitem{33}
Rongjia Yang, Phys. Rev. D {\bf 89}, 063014 (2014).

\bibitem{34}
Ruth Lazkoz, Phys. Rev. D {\bf 86}, 103505 (2012).

\bibitem{35}
V.A. Popov, Phys. Lett. B {\bf 686}, 211-215 (2010).

\bibitem{36}
Peng Wang, Xin-He Meng,  Class. Quant. Grav. {\bf 22,}  283-294 (2005).

\bibitem{37}
J. Weller, A. Albrecht, Phys. Rev. D {\bf 65}, 103512 (2002).

\bibitem{38}
I. Maor et al, Phys. Rev. D {\bf 65}, 123003 (2002).

\bibitem{39}
M. Goliath et al, Astron. Astrophys. {\bf 380}, 6 (2001).

\bibitem{40}
G. Efstathiou, MNRAS {\bf 310}, 842 (1999).

\bibitem{41}
M. Chevallier, D. Polarski, Int. J. Mod. Phys. D {\bf 10}, 213 (2013).

\bibitem{42}
E. V. Linder, Phys. Rev. Lett. {\bf 90,} 091301 (2003).

\bibitem{43}
T. Padmanabhan, T. R. Choudhury, MNRAS {\bf 344}, 823 (2003).

\bibitem{43.1}
E. M. Barboza Jr et al, Phys. Lett. B {\bf666}, 415-419, (2008).

\bibitem{44}
M. Li, Phys. Lett. B {\bf 603}, 1 (2004).

\bibitem{45}
D. Pavon and W. Zimdahl, Phys. Lett. B {\bf 628}, 206 (2005).

\bibitem{46}
R. Horvat, Phys. Rev. D {\bf 70}, 087301 (2004).

\bibitem{47}
X. Zhang and H. Liu, Phys. Lett. B {\bf 659}, 26 (2008).

\bibitem{48}
Chaojun Feng, Phys. Lett. B {\bf 680}, 355-358 (2009).

\bibitem{49}
Masashi Suwa and Takeshi Nihei {\bf 81}, 023519 (2010).

\bibitem{50}
Antonio Pasqua et al, Astrophys. Space. Sci. {\bf 340}, 199-208 (2012).

\bibitem{51}
Changjun Gao et al, Phys. Rev. D {\bf 79}, 043511 (2009).

\bibitem{52}
Jingfei Zhang et al, Phys. Lett. B {\bf 651}, 84-88 (2007).

\bibitem{53}
Jie Ren and Xin-He Meng, Int. J. Mod. Phys. D  {\bf 17}, 2325-2335 (2008).

\bibitem{54}
S. Capozziello, Int. J. Mod. Phys. D {\bf11,} 483 (2002).

\bibitem{55}
S. Capozziello et al, Int. J. Mod. Phys. D {\bf12,} 1969 (2003).

\bibitem{56}
S. M. Carroll et al, Phys. Rev. D {\bf70,} 043528 (2004).

\bibitem{57}
Shin'ichi Nojiri, Sergei D. Odintsov, Phys. Rev. D {\bf68,} 123512 (2003).

\bibitem{58}
L. Amendola, Phys. Rev. D {\bf60,} 043501 (1999).

\bibitem{59}
J. P. Uzan, Phys. Rev. D {\bf59,} 123510 (1999).

\bibitem{60}
T. Chiba, Phys. Rev. D {\bf60,} 083508 (1999).

\bibitem{61}
N. Bartolo, M. Pietroni, Phys. Rev. D {\bf61,} 023518 (2000).

\bibitem{62}
F. Perrotta et al, Phys. Rev. D {\bf61,} 023507 (2000).

\bibitem{63}
V. Sahni and A.A. Starobinsky, Int. J. Mod. Phys. D {\bf 15,} 2105 (2006).

\bibitem{64}
P. Ruiz-Lapuente, Class. Quant. Grav. {\bf 24,} R91 (2007).

\bibitem{A}
S. Nojiri, S. D. Odintsov, Phys. Rev. D {\bf 71}, 123509 (2005).

\bibitem{B}
G. Calcagni et al, Class. Quant. Grav. {\bf 22}, 3977 (2005).

\bibitem{C}
B. M. N. Carter, I.P. Neupane, JCAP {\bf 0606}, 004 (2006).

\bibitem{D}
L. Amendola et al, JCAP {\bf 0612}, 020 (2006).

\bibitem{65}
T. Jacobson, Einstein-aether gravity: a status report, PoS QG-PH: 020 (2007).

\bibitem{66}
Ted Jacobson, Phys. Rev. D {\bf 81}, 101502 (2010).

\bibitem{67}
L. Randall, R. Sundrum, Phys. Rev. Lett. {\bf 83,} 3370 (1999).

\bibitem{68}
L. Randall, R. Sundrum, Phys. Rev. Lett. {\bf 83,} 4690 (1999).

\bibitem{69}
G. R. Davli, G. Gabadadze, M. Porrati, Phys. Lett. B {\bf 485,} 208 (2000).

\bibitem{70}
V. Sahni, Y. Shtanov, JCAP {\bf 0311,} 014 (2003).

\bibitem{71}
V. Sahni, T. D. Saini, A. A. Starobinsky and U. Alam, JETP Lett. {\bf 77}, 201 (2003).

\bibitem{72}
V. Sahni, T. D. Saini and A. A. Starobinsky, MNRAS {\bf 344}, 1057 (2003).

\bibitem{73}
S. Chongchitnan and G. Efstathiou, Phys.Rev. D {\bf 76}, 043508 (2007) [arXiv:0705.1955].

\bibitem{74}
Eric V. Linder, Gen. Rel. Grav. {\bf 40}, 329 (2008)  [arXiv:0704.2064].

\bibitem{75}
V. Gorini, A. Kamenshchik, and U. Moschella, Phys. Rev. D  {\bf 67}, 063509 (2003) [astro-ph/0209395].

\bibitem{76}
P. Wu and H. Yu, Int. J. Mod. Phys. D {\bf 14}, 1873 (2005) [gr-qc/0509036].

\bibitem{77}
Guang Yang, Deng Wang and Xin-He Meng, [arXiv:1602.02552].

\bibitem{78}
X. T. Gao and R. J. Yang, Phys. Lett. B {\bf 687}, 99 (2010) [arXiv:1003.2786].

\bibitem{79}
V. Gorini, A. Kamenshchik, and U. Moschella, Phys. Rev. D  {\bf 67}, 063509 (2003) [astro-ph/0209395].

\bibitem{80}
W. Chakraborty, U. Debnath, and S. Chakraborty, Grav. Cosmol. {\bf 13}, 294 (2007) [arXiv:0711.0079].

\bibitem{81}
S. Li, Y. Ma, Y. Chen, Int. J. Mod. Phys. D  {\bf 18}, 1785 (2009) [arXiv:0809.0617].

\bibitem{82}
L. N. Granda, W. Cardona, and A. Oliveros, [arXiv:0910.0778v1].

\bibitem{83}
J. Zhang, X. Zhang, H. Liu, Phys. Lett. B {\bf 659}, 26 (2008) [arXiv:0705.4145].

\bibitem{84}
C. J. Feng, Phys. Lett. B {\bf 670}, 231 (2008) [arXiv:0809.2502].

\bibitem{85}
H. Wei and R. G. Cai, Phys. Lett. B {\bf 655}, 1 (2007) [arXiv:0707.4526].

\bibitem{86}
R. Yang, J. Qi, and B. Chen, Sci China-Phys Mech Astron {\bf 55}, 1952 (2012).

\bibitem{87}
G. Panotopoulos, Nucl. Phys. B  {\bf 796}, 66 (2008) [arXiv:0712.1177].

\bibitem{88}
R. Myrzakulov, M. Shahalam, JCAP {\bf 10}, 047 (2013) [arXiv:1303.0194].

\bibitem{89}
M. Sami, M. Shahalam, M. Skugoreva, and A. Toporensky, Phys. Rev. D {\bf 86}, 103532 (2012) [arXiv:1207.6691].

\bibitem{90}
S. Ghaffari, A. Sheykhi, and M.H. Dehghani,  Phys. Rev. D {\bf 91}, 023007 (2015).

\bibitem{91}
V. Sahni, A. Shafieloo, and A. A. Starobinsky, Phys. Rev. D {\bf 78}, 103502 (2008) [arXiv:0807.3548].

\bibitem{92}
M. Arabsalmani and V. Sahni, Phys. Rev. D {\bf 86}, 103527 (2012).

\bibitem{93}
Jun Li, Rong-Jia Yang, Bohai Chen, JCAP {\bf 12}, 043 (2014) [arXiv:1406.7514v2].

\bibitem{94}
B. P. Abbott et al, Phys. Rev. Lett. {\bf 116}, 061102 (2016).

\bibitem{95}
A. R. Cooray and D. Huterer, Astrophys. J. {\bf 513}, L95 (1999).

\bibitem{96}
J. Kratochvil et al, JCAP {\bf 0407}, 001 (2004).

\bibitem{97}
C. Blake et al, MNRAS {\bf 418} 1707 (2011).

\bibitem{98}
Tongjie Zhang et al, Adv. Astron. {\bf 2010}, 184284 {2010}.

\bibitem{99}
Xin-He Meng, Xiao-Long Du, Commun. Theor. Phys. {\bf57,} 2 (2012).

\bibitem{100}
M. Visser, Class. Quant. Grav. {\bf 21,} 2603 (2004).

\bibitem{101}
S. Capozziello et al, Phys. Rev. D {\bf 78}, 063504 (2008).

\bibitem{102}
M. P. Dabrowski, Phys. Lett. B. {\bf 625}, 184 (2005).

\bibitem{103}
M. Dunajski and G. Gibbons, Class. Quant. Grav. {\bf 25,} 235012 (2008).

\bibitem{104}
M. Arabsalmani and V. Sahni, Phys. Rev. D {\bf 83}, 043501 (2011).





\end{thebibliography}
\end{document}